\documentclass{aa}  
\usepackage{comment}
\usepackage{siunitx}
\usepackage{subcaption}
\usepackage{nicefrac}
\usepackage{graphicx}
\usepackage{color}
\usepackage{hyperref}
\usepackage{csvsimple}
\usepackage{ulem}
\usepackage[switch]{lineno}

\usepackage{booktabs,wrapfig}
\hypersetup{colorlinks,linkcolor={blue},citecolor={blue}}  

\defcitealias{Casares+2014}{C14}
\defcitealias{Janssens+2023}{J23}
\defcitealias{Rivinius+2024}{R24}

\usepackage{textcomp,gensymb}
\usepackage{caption} 
\usepackage{tabularx}
\usepackage[bottom]{footmisc}
\usepackage{mathtools}
\usepackage[varg]{txfonts}
\begin{document} 

    \title{Ultraviolet spectroscopy reveals a hot and luminous companion to the Be star+black hole candidate MWC 656}
    \titlerunning{UV spectroscopy reveals a hot and luminous companion in the binary MWC 656}

   \author{Johanna M\"uller-Horn\inst{\ref{inst:mpia}, \ref{inst:uni-hd}}\thanks{Corresponding author \email{mueller-horn@mpia.de}} \and
          Varsha Ramachandran\inst{\ref{inst:ari}} \and
          Kareem El-Badry\inst{\ref{inst:caltech},\ref{inst:mpia}} \and
          Andreas A. C. Sander\inst{\ref{inst:ari},\ref{inst:iwr}} \and 
          Julia Bodensteiner\inst{\ref{inst:amsterdam}} \and
          Douglas R. Gies\inst{\ref{inst:chara}} \and
          Ylva G\"otberg\inst{\ref{inst:ista}} \and
          Thomas Rivinius\inst{\ref{inst:eso}} \and
          Tomer Shenar\inst{\ref{inst:telaviv}} \and
          Elisa C. Sch\"osser\inst{\ref{inst:ari}} \and
          Luqian Wang \inst{\ref{inst:yunnan},\ref{inst:chara}} \and
          Allyson Bieryla\inst{\ref{inst:harvard}} \and 
          Lars A. Buchhave\inst{\ref{inst:dtu}} \and
          David W. Latham\inst{\ref{inst:harvard}}
          }

   \institute{
    {Max-Planck-Institut für Astronomie, Königstuhl 17, 69117 Heidelberg, Germany \label{inst:mpia}} 
    \and {Fakultät für Physik und Astronomie, Universität Heidelberg, Im Neuenheimer Feld 226, 69120 Heidelberg, Germany \label{inst:uni-hd}}
    \and {Zentrum für Astronomie der Universität Heidelberg, Astronomisches Rechen-Institut, Mönchhofstr. 12-14, 69120 Heidelberg, Germany \label{inst:ari}}
    \and {Department of Astronomy, California Institute of Technology, Pasadena, CA 91125, USA \label{inst:caltech}}
    \and {Interdisziplin{\"a}res Zentrum f{\"u}r Wissenschaftliches Rechnen, Universit{\"a}t Heidelberg, Im Neuenheimer Feld 225, 69120 Heidelberg, Germany\label{inst:iwr}}
    \and {Anton Pannekoek Institute for Astronomy, University of Amsterdam, Science Park 904, 1098 XH, Amsterdam, The Netherlands\label{inst:amsterdam}}
    \and {Center for High Angular Resolution Astronomy and Department of Physics and Astronomy, Georgia State University, P.O. Box 5060, Atlanta, GA 30302-5060, USA\label{inst:chara}}
    \and {Institute of Science and Technology Austria (ISTA), Am Campus 1, 3400 Klosterneuburg, Austria\label{inst:ista}}
    \and {European Organisation for Astronomical Research in the Southern Hemisphere (ESO), Casilla 19001, Santiago 19, Chile\label{inst:eso}}
    \and {School of Physics and Astronomy, Tel Aviv University, Tel Aviv 6997801, Israel\label{inst:telaviv}}
    \and {Yunnan Observatories, Chinese Academy of Sciences (CAS), Kunming 650216, Yunnan, People's Republic of China\label{inst:yunnan}}
    \and {Center for Astrophysics ${\rm \mid}$ Harvard {\rm \&} Smithsonian, 60 Garden Street, Cambridge, MA 02138, USA \label{inst:harvard}}
    \and {DTU Space,  Technical University of Denmark, Elektrovej 328, DK-2800 Kgs. Lyngby, Denmark \label{inst:dtu}}}

   \date{Received 04 November 2025 / Accepted 28 January 2026}

 
  \abstract
    {The Galactic Be star binary MWC~656 was long considered the only known Be star+black hole (BH) system, making it a critical benchmark for models of massive binary evolution and for the expected X-ray emission of Be+BH binaries. However, recent dynamical measurements cast doubt on the presence of a BH companion. We present new multi-epoch ultraviolet spectroscopy from the \textit{Hubble} Space Telescope, combined with high-resolution optical spectra, to reassess the nature of the companion. The far-ultraviolet spectra reveal high-ionisation features -- including prominent N~\textsc{v} and He~\textsc{ii} lines -- which are absent in the spectra of normal Be stars and are indicative of a hot, luminous companion. Spectral modelling shows that these features cannot originate from the Be star or from an accretion disc around a compact object. Instead, we find that the data are best explained by a hot ($T_\mathrm{eff}\approx85\,$kK), compact, hydrogen-deficient star with strong wind signatures, consistent with an intermediate-mass stripped star. Our revised orbital solution and composite spectroscopic modelling yield a companion mass of $M_2 = 1.48^{+0.55}_{-0.46}\,\mathrm{M}_\odot$, definitively ruling out a BH and disfavouring a white dwarf. MWC~656 thus joins the growing class of Be + stripped star binaries. The system's unusual properties -- including a high companion temperature and wind strength -- extend the known parameter space of such binaries. The continued absence of confirmed OBe+BH binaries in the Galaxy highlights a growing tension with population synthesis models.} 

   \maketitle

\section{Introduction}

The population of black holes (BHs) in binaries with massive stellar companions remains poorly constrained. While numerous high-mass X-ray binaries are known \citep{Fortin+2023}, only a few host dynamically confirmed BH accretors, such as Cyg~X-1, LMC~X-1, and M33~X-7 \citep[e.g.][]{Ramachandran+2025,Ramachandran2022, Orosz+2007, Orosz+2009}. Even more elusive are non-accreting BHs in detached systems, with only a single such O+BH binary currently identified in the Large Magellanic Cloud \citep[VFTS243;][]{Shenar+2022} and two Galactic candidate systems \citep[HD~130298 and ALS~8814;][An et al. in prep., but see also \citealt{El-Badry+2025}]{Mahy+2022}.

This observational scarcity contrasts with predictions from population synthesis models, which suggest that BH companions to late O- and early B-type stars should be relatively common. Models predict that up to a few percent of such stars could host BHs in detached configurations \citep{Langer+2020, Xu+2025, Schürmann+2025}. These OB+BH binaries are expected to represent a natural intermediate phase in the evolution of massive binaries en route to double compact object formation, a subset of which may merge as gravitational-wave sources.

In the stable mass-transfer formation channel, the primary star is stripped of much of its envelope before collapsing into a BH, whereas the secondary star may accrete mass and angular momentum \citep[e.g.][]{Mapelli2021, Marchant_Bodensteiner2024}. This process often results in a rapidly rotating secondary, potentially manifesting as a Be or Oe star \citep{deMink+2013}. Given that many Be stars are believed to be spun-up mass gainers \citep{Pols+1991,Hastings+2021}, the search for BHs in Be-star binaries is particularly compelling. Whether the current lack of such detections reflects observational biases or shortcomings in binary evolution models remains a key open question in the study of massive stars.

A recurring challenge in identifying OB+BH systems lies in the misinterpretation of their components. Several recent OB+BH candidates \citep{Liu+2019,Rivinius+2020} have ultimately been reclassified as luminous binaries with bloated stripped star companions \citep{Shenar+2020,El-Badry_Quataert2020,Bodensteiner+2020}. Systems with large mass functions and apparently unseen companions were later shown to contain a hot, stripped star and a rapidly rotating early-type companion \citep[e.g.][]{El-Badry_Quataert2020b,Mueller-Horn+2025}. These cases underscore the difficulty of securely identifying quiescent BHs and the need for high-quality multi-wavelength observations.

MWC~656 was the first candidate Galactic Be+BH binary and was reported by \citet[][hereafter \citetalias{Casares+2014}]{Casares+2014}. The detection was based on optical spectroscopy that revealed He~\textsc{ii} 4686\,Å emission, which was interpreted as arising from an accretion disc around an unseen compact object. Using the radial velocity (RV) curves of both the Be star and the He~\textsc{ii} emission line, \citetalias{Casares+2014} derived a modest eccentricity ($e = 0.10 \pm 0.04$) and a mass ratio ($q$) of $M_\mathrm{Be}/M_2 \approx 2.4$, implying a companion mass of 3.8--6.9\,M$_\odot$ for an assumed Be star mass of 10--16\,M$_\odot$. The absence of spectroscopic features from a luminous companion supported the interpretation of a BH. Additional evidence came from the faint X-ray luminosity of the system \citep[$L_X \simeq 3 \times 10^{30} \,\text{erg}\, \mathrm{s}^{-1}$;][]{Ribo+2017} and its location in the radio/X-ray luminosity plane, which is consistent with the properties of quiescent BH binaries \citep{Dzib+2015,Ribo+2017}. The system was also suggested as a candidate $\gamma$-ray binary \citep{Lucarelli+2010}, although subsequent analysis showed that the detected emission likely originates from a background quasar \citep{Alexander_McSwain2015}. The Be star itself has been classified as a B1.5--2\,III star \citepalias{Casares+2014}, with an effective temperature ($T_{\rm eff}$) of $19 \pm 3$\,kK and a mass of roughly 10\,M$_\odot$ as derived by \citet{Williams+2010}. Its rotational velocity was measured at $v \sin i \approx 313$\,km\,s$^{-1}$ \citep{Zamanov+2021}.

However, the interpretation of MWC~656 as a Be+BH binary has been increasingly questioned. \citet[][hereafter \citetalias{Rivinius+2024}]{Rivinius+2024} re-analysed the system using high-resolution optical spectroscopy and found lower RV amplitudes for the Be star ($K_{\rm Be} = 10$--15\,km\,s$^{-1}$ based on \ion{He}{i}\,$\lambda$6678 absorption), compared to the $K_{\rm Be} = 32.0 \pm 5.3$\,km\,s$^{-1}$ derived by \citetalias{Casares+2014} using Fe~\textsc{ii} emission. This discrepancy led to a revised, much lower companion mass that is inconsistent with a BH and instead suggests a hot subdwarf companion. Independent observations by \citet[][hereafter \citetalias{Janssens+2023}]{Janssens+2023} supported this view. They inferred a mass ratio $q \approx 8.3$ and favoured a hot subdwarf, white dwarf, or neutron star companion. The different RV amplitudes reported in these studies likely arise from the choice of spectral tracers. While \ion{He}{i} absorption lines trace the photosphere of the Be star, \ion{Fe}{ii} emission originates in the circumstellar disc and can vary with tidal distortion or disc asymmetries \citep{Panoglou+2016}, possibly biassing velocity measurements (see also Sect.~\ref{sec:binary_orbit}).

The uncertain nature of the companion in MWC~656 motivates further investigation. Ultraviolet (UV) spectroscopy offers a powerful means to resolve this ambiguity: stripped helium stars emit most of their radiation in the UV, resulting in a more favourable flux ratio compared to optical wavelengths. This facilitates their detection in composite spectra, as demonstrated by the discovery of ten Galactic Be+subdwarf binaries by \citet{Wang+2021} using UV spectroscopy, and offers the potential to constrain the wind properties and ionising flux of the companion. UV observations also yield improved constraints on the properties of the Be star, in particular its mass, luminosity, and wind parameters, since optical spectra can be contaminated by emission from the circumstellar disc. A detailed characterisation of the Be star is still lacking, and disentangling its spectral contribution is essential to correctly inferring the companion mass. 

In this study we combined new high-resolution optical spectroscopy with UV spectra obtained using the \textit{Hubble} Space Telescope (HST). The optical data enabled a refined orbital solution and updated constraints on the Be star’s parameters. The HST far-ultraviolet (FUV) spectroscopy allowed us to search directly for the signature of a hot, luminous companion. This paper is organised as follows: In Sect.~\ref{sec:observations} we describe the observations and data reduction. Section~\ref{sec:binary_orbit_and_disentangling} presents the revised orbital solution, Sect.~\ref{sec:BeStar} details the analysis of the Be star, and Sect.~\ref{sec:UV} examines the UV spectra for evidence of a companion. In Sect.~\ref{sec:model} we jointly model UV and optical spectra with a composite binary model. In Sect.~\ref{sec:companion} we discuss the nature of the companion. We place MWC~656 in the context of known post-interaction binaries in Sect.~\ref{sec:discussion} before summarising our conclusions in Sect.~\ref{sec:conclusion}.

\section{Observations and data reduction}
\label{sec:observations}

\subsection{Optical spectroscopy}

We obtained nine new optical spectra of MWC~656 using the Tillinghast Reflector Echelle Spectrograph (TRES; \citealt{Furesz2008}) on the 1.5\,m Tillinghast Reflector telescope at the \textit{Fred Lawrence Whipple} Observatory, Mount Hopkins, Arizona. TRES has a resolving power ($R$) of $\simeq 44000$ and covers a wavelength range from $\sim$3900 to 9100\,Å. The spectra were reduced following the procedure described in \citet{Buchhave+2010}, including bias subtraction, flat-field correction, and wavelength calibration.

In addition, we analysed five spectra obtained with the High Resolution Echelle Spectrometer (HIRES; \citealt{Vogt+1994}) on the Keck I telescope at W. M. Keck Observatory, Mauna Kea. HIRES provides a spectral resolving power of $\simeq 55000$. These data were reduced using the standard California Planet Survey pipeline (CPS; \citealt{Howard+2010}), and the extracted spectra span three wavelength ranges: $\sim$ 3700--4800\,Å, 5000--6400\,Å, and 6550--8000\,Å. For every TRES and HIRES spectrum, we merged overlapping echelle orders to create continuous spectra, using linearly weighted averages in the overlapping regions.

We also retrieved archival high-resolution spectra from multiple facilities. These include three epochs of observations with the Echelle SpectroPolarimetric Device
for the Observation of Stars (ESPaDOnS, $R \approx 68000$; \citealt{Donati+2003}) at the Canada-France-Hawaii Telescope (CFHT), covering $\approx 3700$--10000\,Å. Each ESPaDOnS spectrum consists of eight co-added exposures taken on a single night to increase the signal-to-noise ratio (S/N). Furthermore, we retrieved 12 archival spectra obtained with the ARC Echelle Spectrograph (ARCES; \citealt{Wang+2003}) at Apache Point Observatory. ARCES offers a resolving power of $\simeq 31500$, with coverage from 3500 to 10500\,Å.

We initially considered the 34 archival spectra taken with the Fibre-fed RObotic Dual-beam Optical Spectrograph (FRODOspec; \citealt{Morales-Rueda+2004}), which were also used by \citetalias{Casares+2014}, \citetalias{Janssens+2023} and \citetalias{Rivinius+2024}. However, as noted by \citetalias{Janssens+2023}, these data are affected by a non-linear wavelength calibration issue. Our own inspection of interstellar \ion{Ca}{ii} H and K absorption features confirms wavelength offsets of the order of $5\,\AA$, which decrease towards longer wavelengths. \citetalias{Janssens+2023} mitigated this by using selected wavelength regions and applying local corrections. Since the affected spectra are not essential to our analysis and residual distortions could not be fully corrected by a simple velocity shift, we excluded them to avoid introducing systematic uncertainties in our RV measurements.

All 29 retained spectra were individually continuum-normalised using cubic spline fits to line-free regions. The spline anchor points were manually selected in wavelength intervals free of strong absorption or emission.
We applied barycentric corrections to all spectra and empirically estimated the S/N value from the pixel-to-pixel variation in continuum regions. A summary of all optical observations, including dates and typical S/N values, is provided in Table~\ref{tab:obs_overview}.

\subsection{UV spectroscopy}

We obtained multi-epoch UV spectroscopy of MWC~656 with the HST, combining observations from two programmes (programme \texttt{17074} PI: V. Ramachandran, programme \texttt{17202} PI: K. El-Badry). The dataset includes:

\begin{itemize}
    \item A FUV spectrum obtained with the Cosmic Origins Spectrograph (COS) using segment B of the FUV channel with the G130M grating (central wavelength 1096\,Å), covering $\sim$940--1090\,Å and with a resolving power ($R$) of $\sim 5000$--$12000$.
    \item Four FUV spectra obtained with the Space Telescope Imaging Spectrograph (STIS) using the FUV-MAMA detector and the E140M echelle grating, providing a resolving power of $45800$ over 1150--1700\,Å.
    \item A near-UV spectrum obtained with STIS/NUV-MAMA using the E230M echelle grating, covering 1600--3100\,Å with a resolving power of $30000$.
\end{itemize}

These HST observations span modified Julian dates (MJD) 59920 to 60347. Three of the STIS/FUV spectra were obtained over a 10-week interval, with the fourth taken approximately one year later. All observations are listed in Table~\ref{tab:obs_overview}.

We retrieved fully reduced, flux- and wavelength-calibrated spectra from the 
\href{http://dx.doi.org/10.17909/cfms-dg96}{MAST} archive. COS data were processed using the standard \texttt{CalCOS} pipeline \citep{Soderblom2022}, while STIS data were reduced with \texttt{CalSTIS} \citep{Sohn2019}. For the STIS echelle spectra, we merged the individual orders onto a 1D wavelength grid uniformly spaced in logarithmic wavelength.

\section{Binary orbit and spectral disentangling}
\label{sec:binary_orbit_and_disentangling}

\begin{figure*}[t]
    \centering
    \begin{minipage}[t]{0.33\textwidth}
        \centering
        \includegraphics[width=\textwidth]{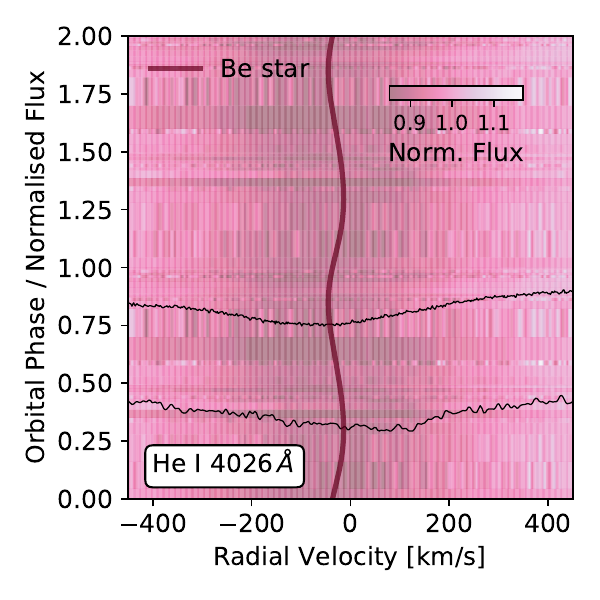}
    \end{minipage} %
    \begin{minipage}[t]{0.33\textwidth}
        \centering
        \includegraphics[width=\textwidth]{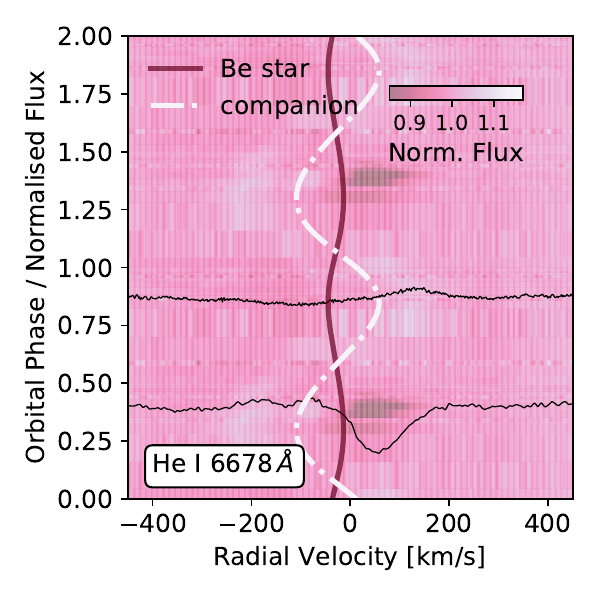}
    \end{minipage} %
    \begin{minipage}[t]{0.33\textwidth}
        \centering
        \includegraphics[width=\textwidth]{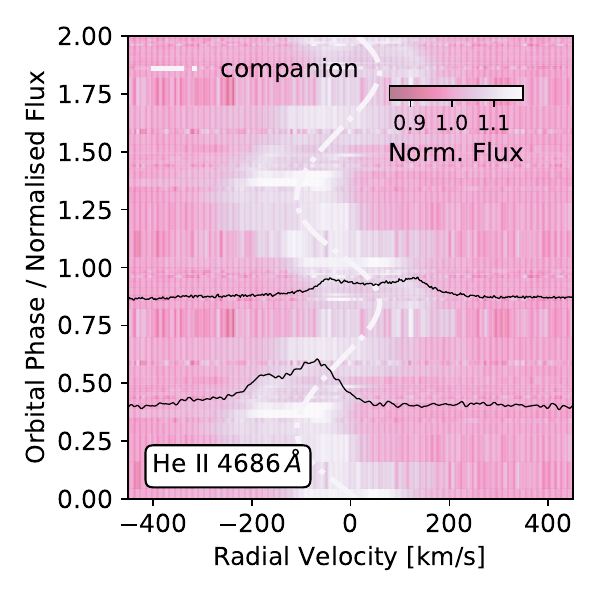}
    \end{minipage}
    \caption{Dynamical spectra of MWC~656 shown over two orbital cycles. From left to right, the panels display the \ion{He}{i\,$\lambda4026$}, \ion{He}{i\,$\lambda6678$}, and \ion{He}{ii\,$\lambda4686$} lines. Two epoch spectra near quadrature phases are overplotted in black to further illustrate line profile variations. The orbital RV curves of the Be star (pink) and the companion (white), derived from the best-fit orbit model, are overlaid and trace the associated absorption and emission features.} 
    \label{fig:doppler_tomography}
\end{figure*}

The optical spectrum of MWC~656 is dominated by its Be star component. Characteristic features include strong, double-peaked emission in hydrogen Balmer lines and shallow, rotationally broadened absorption lines, most prominently \ion{He}{i}. These spectral signatures are typical of rapidly rotating Be stars surrounded by a circumstellar decretion disc. The disc also produces a series of double-peaked \ion{Fe}{ii} emission lines \citep[e.g. \ion{Fe}{ii\,$\lambda\lambda$4583, 5198};][]{Rivinius+2013}, clearly visible in the spectra. The RV variability of the Be star is modest in amplitude, as shown by the Doppler shift of the \ion{He}{i\,$\lambda4026$} absorption line (Fig.~\ref{fig:doppler_tomography}, left panel).

The clearest spectral signature of the companion star is a strong \ion{He}{ii\,$\lambda$4686} emission line. The line profile is slightly double-peaked and exhibits significant RV variability. Its motion is in anti-phase with the Be star and shows a much larger RV amplitude (Fig.~\ref{fig:doppler_tomography}, right panel), supporting its association with the binary companion.
Closer inspection of the \ion{He}{i} spectral lines, particularly at redder wavelengths, reveals subtle signs of the companion as well: A weak emission component moving in anti-phase with the Be star is especially apparent in the \ion{He}{i\,$\lambda$6678} line, as noted by \citetalias{Rivinius+2024}. In addition, this line shows transient narrow absorption features, reminiscent of shell lines seen in other Be+subdwarf binaries \citep[e.g. $\phi$~Per;][]{Poeckert1981,Stefl+2000}. This is illustrated in the middle panel of Fig.~\ref{fig:doppler_tomography}.

\subsection{Radial velocities}
To measure the RVs of the Be star, we cross-correlated the observed spectra with a synthetic template spectrum. We used a model atmosphere generated with the Potsdam Wolf-Rayet (PoWR) code for OB-type stars \citep{Graefener+2002, Hamann_Graefener2003, Sander+2015}. This code iteratively solves the radiative transfer equation for a spherically expanding atmosphere and statistical equilibrium equations. PoWR is a state-of-the-art code in the field of analysing hot stars with winds (see e.g. \citealt{Sander+2024} for a recent comparison) and has been used successfully for the spectral analysis of Be binaries with a stripped star companion \citep[e.g.][]{Ramachandran+2023,Ramachandran+2024}. For the cross-correlation template, we adopted a model with $T_\mathrm{eff} = 19\,\mathrm{kK}$, $\log g = 3.0$, and solar metallicity from the OB-type Milky Way grid \citep{Hainich+2019}.

The cross-correlation was carried out simultaneously on multiple He~\textsc{i} absorption lines (\ion{He}{i\,$\lambda\lambda3820,\,4009,\,4026,\,4144$}), located primarily in the blue part of the spectra. These lines were chosen because they are less affected by contamination from emission components originating in the companion or the disc of the Be star.

For the companion star, the \ion{He}{ii\,$\lambda$4686} emission line is the only clear and reliable tracer of orbital motion in the optical spectrum because of the absence of detectable absorption features. We determined its RVs through iterative cross-correlation centred on the He~\textsc{ii} emission region. To this end, we used the spectrum with the highest S/N as the initial template to derive relative RVs. In subsequent iterations, a new cross-correlation template was constructed by co-adding the spectra after shifting them to the rest frame according to the derived velocities. This method is based on relative shifts and does not constrain the companion's systemic velocity, which we fitted instead during orbital analysis.

The RVs determined for both components are listed in Table~\ref{tab:obs_overview}.
RVs were measured by fitting a parabola to the peak of the cross-correlation function. The uncertainty in each RV was estimated following the prescription from \cite{Zucker2003}. 
This approach yields an analytical estimate of the formal uncertainty based on the local shape of the cross-correlation function peak. The quoted uncertainties represent lower bounds because they do not account for systematic effects such as template mismatch.

\subsection{Orbital analysis}
\label{sec:binary_orbit}

\begin{figure}[t]
    \centering
    \includegraphics[width=\columnwidth]{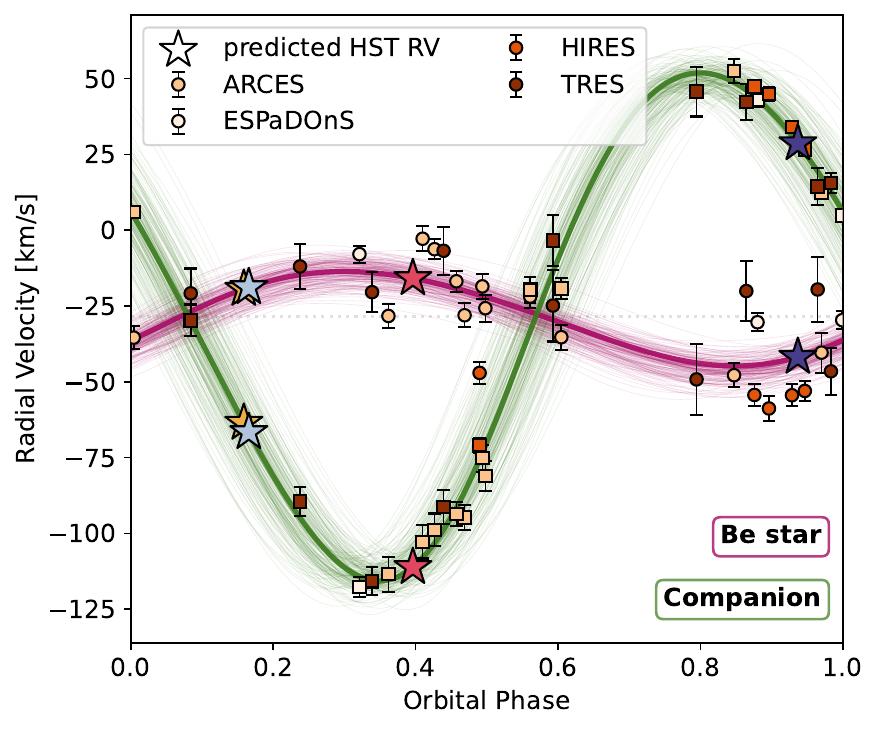}
    \caption{Phase-folded RV curves of the Be star and its companion in the MWC~656 binary system, using the best-fit orbital period. RV measurements and uncertainties of the Be star (circles) and the companion (squares) are shown with colour-coded markers corresponding to the instrument used.
    Maximum-likelihood orbital models derived from RV fitting are overplotted for the Be star (pink) and the companion (green) with thick lines. Thin lines show orbital models for random samples from the posterior to illustrate uncertainties in the inferred parameters. The grey dotted line shows the inferred systemic velocity. Star symbols indicate the predicted orbital phases and velocities of the four HST/STIS FUV observations based on the fitted orbit.} 
    \label{fig:orbit_fit}
\end{figure}

\begin{table}[t]
\setlength{\extrarowheight}{2.5pt}
\centering
\caption{Orbital parameters and uncertainties.}
\begin{tabular}{l c c}
\hline
\hline
Orbital period & $P$ [d] & {$59.037 \pm 0.014$} \\ 
Eccentricity & $e$ & {$0.071\pm0.028$} \\ 
Argument of periastron & $\omega$ [rad]& {$4.19\pm0.45$} \\ 
Epoch of periastron & $T_\mathrm{peri}$ [MJD] & {$51541.0\pm3.7$} \\ 
Barycentric velocity & $v_z$ [km s$^{-1}$] & {$-29.4 \pm 2.3$} \\ 
Relative RV offset & $v_{\mathrm{rel.}, 2}$ [km s$^{-1}$] & {$-32.6 \pm 1.8$} \\
RV semi-amplitude &  & \\ 
-- Be star  & $K_{\mathrm{Be}}$ [km s$^{-1}$] & {$16.5 \pm 3.1$} \\ 
-- Companion  & $K_{\mathrm{2}}$ [km s$^{-1}$] & {$83.4\pm2.1$} \\  
Dynamic mass&  & \\
-- Be star  & $M_{\mathrm{Be}} \sin i ^3$ [M$_\odot$] & {$5.13 \pm 0.47$} \\ 
-- Companion  & $M_{\mathrm{2}} \sin i ^3$ [M$_\odot$] & {$1.03 \pm 0.26$} \\ 
Dynamic mass ratio & $q_\mathrm{dyn}=\nicefrac{K_\mathrm{2}}{K_\mathrm{Be}}$ & {$5.25^{+0.91}_{-0.98}$} \\ 
RV jitter & & \\
-- Be star & $s_\mathrm{Be}$ [km s$^{-1}$] & {$10.4\pm1.9$} \\ 
-- Companion & $s_2$ [km s$^{-1}$] & {$4.7\pm1.3$} \\
\hline
\hline
\end{tabular}
\label{tab:orbit_parameters}
\setlength{\extrarowheight}{0pt}
\end{table}

Previous orbital solutions for the MWC~656 system were presented by \citetalias{Casares+2014} and \citetalias{Janssens+2023}. However, given the addition of new high-resolution spectroscopy and revised RV measurements, we opted to perform an independent orbital analysis for consistency.

We inferred the orbital parameters by simultaneously fitting the RV curves of both the Be star and its companion using a nested sampling framework. Specifically, we used the \texttt{MLFriends} algorithm \citep{Buchner2016, Buchner2019}, implemented in the package \href{https://johannesbuchner.github.io/UltraNest/}{UltraNest} \citep{Buchner2021}, to sample posterior probability distributions.

The orbit of a double-lined spectroscopic binary can be described by seven parameters: the RV semi-amplitudes of the two components ($K_\mathrm{Be}$, $K_2$), the orbital period ($P$), eccentricity ($e$), argument of periastron ($\omega$), systemic velocity ($v_z$), and a phase parameter ($\tau$), which we defined relative to a fixed reference time ($T_\mathrm{ref} = \text{MJD\,} 51544$) as 
$\tau = \left(\frac{T_\mathrm{ref} - T_\mathrm{peri}}{P}\right) \bmod 1\,.$
This parameterisation follows the implementation used in \citet{Blunt+2020}. Since the RVs of the companion were determined relative to a template and lack an absolute zero point, we introduced an additional free parameter ($v_{\mathrm{rel.}, 2}$), which represents a constant velocity offset between the measured relative companion RVs and the barycentric velocity of the system. To account for potentially underestimated uncertainties, we also fitted RV jitter terms ($s_\mathrm{Be},\,s_2$) that are added in quadrature to the RV uncertainties determined from the cross-correlation function peak. 
We adopted uniform priors for all model parameters. The prior ranges were as follows: $P \in [1, 100]$\,d; $K_\mathrm{Be}, K_2 \in [0, 100]$\,km\,s$^{-1}$; $\tau, e \in [0, 1)$; $\omega \in [0, 2\pi]$; $v_z, v_{z,2} \in [-100, 100]$\,km\,s$^{-1}$; and $s_\mathrm{Be}, s_{2} \in [0, 30]$\,km\,s$^{-1}$.

The best-fit orbital solution is shown in Fig.~\ref{fig:orbit_fit}, and the inferred parameters are listed in Table~\ref{tab:orbit_parameters}. We find an orbital period of $P = 59.037 \pm 0.014$\,d and a mildly eccentric orbit ($e = 0.071 \pm 0.028$), consistent with previous estimates. The inferred RV semi-amplitudes are $K_\mathrm{Be} = 16.5 \pm 3.1$\,km\,s$^{-1}$ and $K_2 = 83.4 \pm 2.1$\,km\,s$^{-1}$, corresponding to a dynamical mass ratio $q_\mathrm{dyn} = M_\mathrm{Be}/M_2 = K_2/K_\mathrm{Be} = 5.25^{+0.91}_{-0.98}$. These values imply minimum dynamical masses of $M_\mathrm{Be} \sin^3 i = 5.13 \pm 0.47\,M_\odot$ and $M_{2} \sin^3 i = 1.03 \pm 0.26\,M_\odot$.

Our dynamical mass ratio is lower than the $q \approx 8.3 \pm 2.1$ reported by \citetalias{Janssens+2023}, who measured $K_\mathrm{Be} \approx 11.5$\,km\,s$^{-1}$ and $K_2 \approx 93$\,km\,s$^{-1}$. The discrepancy may arise from the smaller and lower-S/N dataset used in that study (18 HERMES spectra compared to 29 spectra used here). 
\citetalias{Casares+2014} reported a similar companion semi-amplitude ($K_2 = 78.1 \pm 3.2$\,km\,s$^{-1}$) but a substantially higher Be star amplitude ($K_\mathrm{Be} = 32.0 \pm 5.3$\,km\,s$^{-1}$), possibly due to the inclusion of FRODOspec data, which we find suffer from unreliable wavelength calibration, or because \ion{Fe}{ii} emission lines were used as RV tracers. As noted by \citetalias{Rivinius+2024}, the double-peaked emission lines from the Be star’s disc vary in peak separation, likely due to tidal distortion of the disc \citep{Panoglou+2016, Rubio+2025}, making them unreliable for measuring RV amplitudes.

\subsection{Spectral disentangling}
\label{sec:disentangling}
\begin{figure*}[ht]
    \centering
    \begin{minipage}[t]{0.33\textwidth}
        \centering
        \includegraphics[width=\textwidth]{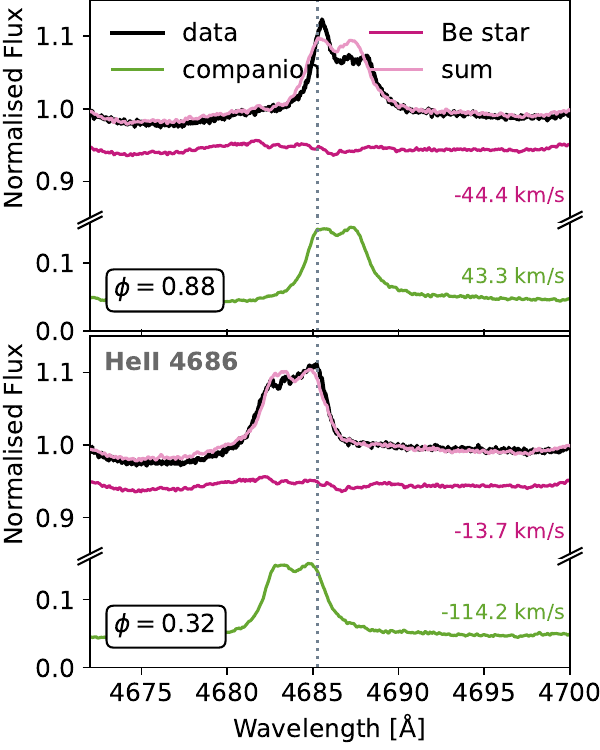}
    \end{minipage} %
    \begin{minipage}[t]{0.33\textwidth}
        \centering
        \includegraphics[width=\textwidth]{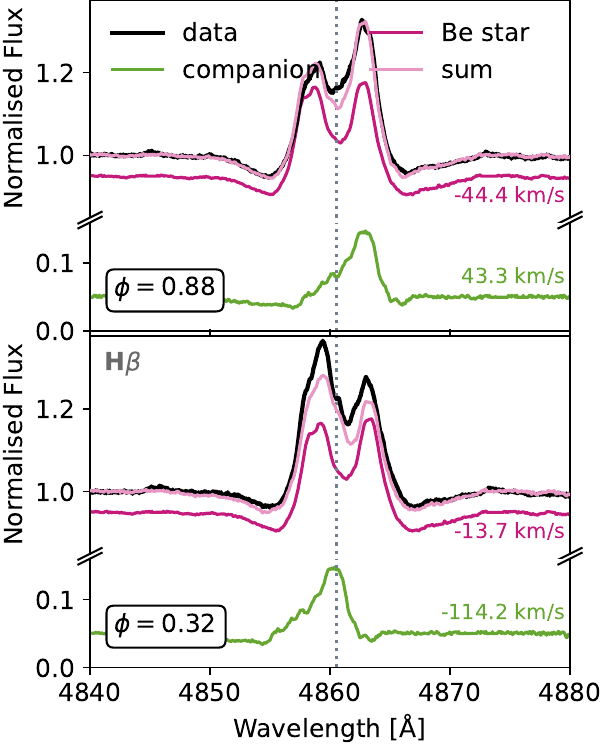}
    \end{minipage} %
    \begin{minipage}[t]{0.33\textwidth}
        \centering
        \includegraphics[width=\textwidth]{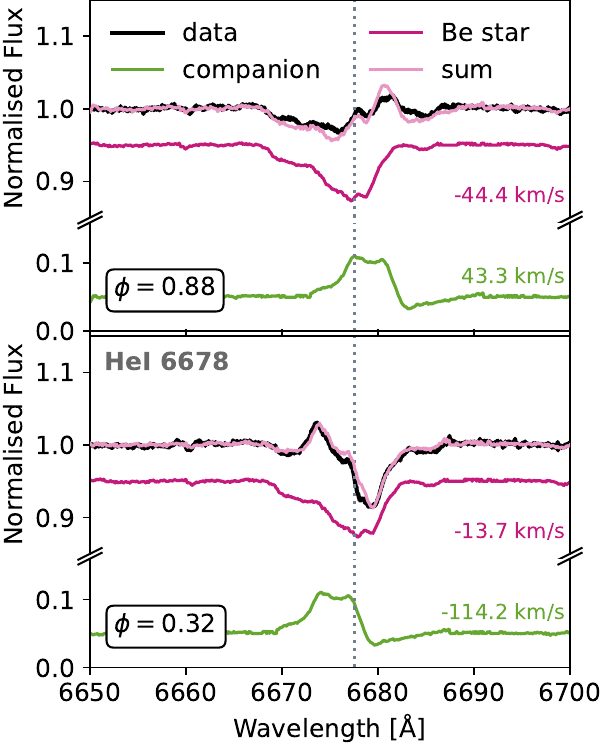}
    \end{minipage}
    \caption{Disentangled spectra of MWC~656 near quadrature, showing spectral regions where the companion exhibits non-flat continuum features. From left to right, the panels display the \ion{He}{ii\,$\lambda$4686} emission line, Balmer H$\beta$, and \ion{He}{i\,$\lambda$6678}. Observed spectra near quadrature phases are plotted in black (top and bottom rows). The individual Doppler-shifted components of the Be star (pink) and the companion (green), obtained from spectral disentangling, are overplotted together with their summed model (light pink).} 
    \label{fig:disentangling}
\end{figure*}

We attempted spectral disentangling to separate the composite spectra into individual components and search for hidden features of the companion. For this, we applied the shift-and-add technique \citep{Marchenko+1998} using the algorithm of \citet{Gonzalez_Levato2006} and the implementation by T. Shenar\footnote{\url{https://github.com/TomerShenar/Disentangling_Shift_And_Add}} \citep{Shenar+2020, Shenar+2022b}. The method starts out with a flat spectrum for the companion and a stacked spectrum created from the Doppler-shifted, co-added observations for the primary. Then it refines both components in an iterative process.
We adopted the orbital parameters from Sect.~\ref{sec:binary_orbit} and disentangled the normalised spectra across the shared wavelength range from 4000\,Å to 6700\,Å. To avoid introducing spurious features, we constrained the disentangled spectra to remain below the continuum in regions without emission.

The resulting companion spectrum is largely featureless, with no clear absorption lines detected. Figure~\ref{fig:disentangling} shows the sum of the disentangled spectra overplotted on observed spectra at quadrature, for three selected regions where the companion spectrum shows non-flat features. As expected, the \ion{He}{ii\,$\lambda$4686} emission line is clearly associated with the companion (left panel). The \ion{He}{i} lines and H$\beta$ are also separated successfully, with the companion showing broad, double-peaked emission profiles that move in anti-phase with the Be star (middle and right panels).
However, interpretation of the \ion{He}{i} and Balmer-line regions is complicated by potential variability in emission components associated with the Be-star disc and/or the companion. Such variability would violate the key assumption of spectral disentangling, that the individual component spectra remain time-invariant aside from Doppler shifts, and may introduce artefacts into the extracted component spectra, including distortions of the Be-star absorption profiles.

The absence of absorption features in the companion spectrum may be due to several factors: the companion could be a dark remnant, with all observed emission originating from a surrounding accretion disc; alternatively, it could be a luminous object with shallow or sparse absorption features or with a light contribution too weak to detect them. In Fig.~\ref{fig:disentangling}, we adopted a continuum flux ratio of 0.95 (Be star) to 0.05 (companion) for visualisation purposes, but the flux ratio itself is unconstrained by disentangling and must be inferred independently.

\section{Stellar parameters of the Be star}
\label{sec:BeStar}
Reliable stellar parameters of the Be star are essential for constraining the nature of the companion. Combined with the dynamical mass ratio, they provide the basis for estimating the companion mass and allow us to model the Be star's UV contribution, necessary for identifying features from the companion.  

We first analysed the Be star under the assumption that it dominates the observed light, that is, a `single-star' approach treating the companion as dark. This provides a first-order characterisation of the Be star's properties. If the companion contributes significantly, the derived luminosity and radius would be slight overestimates (though Sect.~\ref{sec:model} shows such corrections are small). This single-star analysis establishes the foundation for identifying discrepant spectral features in Sect.~\ref{sec:UV} and for composite modelling in Sect.~\ref{sec:model}.  

We estimated the fundamental parameters of the Be star by simultaneously fitting diagnostic lines in the optical and UV spectra with synthetic models. We used the highest S/N normalised optical spectrum shifted to the Be star rest frame and the $\Phi=0.4$ UV spectrum, and compared them to a grid of synthetic models from the non-local thermodynamic equilibrium PoWR model atmosphere code \citep{Graefener+2002, Hamann_Graefener2003, Sander+2015}, convolved with instrumental broadening kernels. As a starting point, we selected models from the OB-type Milky Way grid \citep{Hainich+2019} with solar metallicity and then computed additional models varying $T_\mathrm{eff}$ and $\log g$ to refine the fit.
The effective temperature of the Be star was predominantly constrained using the \ion{Si}{iii\,$\lambda\lambda4553-68-75$}/\ion{Si}{ii\,$\lambda\lambda4128-31$} and \ion{He}{i\,$\lambda4471$}/\ion{Mg}{ii\,$\lambda4481$} line ratios, while surface gravity was estimated from the pressure-broadened wings of the Balmer lines. We primarily used the H$\gamma$, H$\delta$, and H$\epsilon$ lines, which are less affected by emission from the circumstellar disc, and also considered the outer wings of H$\beta$, explicitly excluding the line cores in all cases.

To estimate projected rotational velocity ($v\sin i$) and macroturbulent velocity ($v_\mathrm{mac}$), we applied the Fourier transform and the goodness-of-fit technique implemented in the \texttt{iacob-broad} tool \citep{SimonDiaz_Herrero2014}. We used He~\textsc{i} absorption lines in the blue part of the optical spectra (e.g. \ion{He}{i\,$\lambda\lambda$4026, 4144, 4388}). We determined $v\sin i = 298 \pm 50$\,km\,s$^{-1}$ and $v_\mathrm{mac} = 147 \pm 30$\,km\,s$^{-1}$ by averaging the results over individual lines, with the standard deviation representing the uncertainties (we note that the actual macroturbulent broadening could be lower, given the degeneracy with $v\sin i$). These values were then used to convolve the synthetic spectra for a direct comparison with the observations.

We modelled the spectral energy distribution (SED) using the flux-calibrated UV spectra and archival photometry in the U, B, V, and \textit{Gaia} G bands \citep{Jaschek_Egret1982, Fabricius+2002, GaiaDR3}. Near-infrared bands \citep[2MASS $J$, $H$, and $K_s$;][]{Cutri+2003} were not included in the fit to avoid contamination by thermal emission from the circumstellar disc of the Be star, causing an infrared excess \citep{Williams+2010}. 
We adopted the extinction law as prescribed by \citet{Fitzpatrick1999}. We simultaneously optimised the colour excess ($E(B-V)$), the total-to-selective extinction ratio ($R_V$), and a luminosity ($\log L_\ast$) scaling factor. The model flux was scaled using a fixed distance modulus of 11.47, which corresponds to 1.97\,kpc and is based on the \textit{Gaia} Data Release 3 parallax measurement ($\varpi = 0.4860\pm0.0185$\,mas; \citealt{GaiaDR3, Lindegren+2020})\footnote{We corrected the parallax measurement for the \textit{Gaia} parallax zero point of $-0.023$\,mas \citep{Lindegren+2020} but note that the resulting changes in distance modulus by 0.1\,mag and system luminosity $\log L$ by $\sim0.04$ are negligible compared to other uncertainties in the stellar parameters.}. The optimal fit parameters were determined through iterative $\chi^2$ minimisation, ensuring that the model SED accurately matched both the observed spectrum and multi-band photometry.

The best-fit model yields $T_\ast = 21 \pm 1$\,kK, $\log g_\ast = 3.4 \pm 0.1$, and $\log L_\ast/\mathrm{L}_\odot = 3.93$, corresponding to a radius of $R_\ast = 7.1\,R_\odot$. The uncertainties quoted reflect the quality of the fit and the resolution of the model grid. 
While the temperature is considered reliable and consistent with the spectral type B1.5--2\,III, the derived surface gravity may be underestimated. Rapid rotation significantly changes the Be star's spherical structure, causing it to flatten at the poles and bulge at the equator. This oblateness means that the stellar radius and surface gravity are no longer uniform across the stellar surface, with both differing from pole to equator. When considering the effect of the Be star's rapid rotation ($v\sin i = 300$ km\,s$^{-1}$), the dynamically corrected surface gravity \citep[$\log g_{\text{true}} \coloneqq \log (g_\ast +(\varv \sin i)^{2} /R_\ast)$;][]{Repolust+2004} is $\log g_{\text{true}} = 3.64 \pm 0.10$. 
This `true $\log g$' provides a more accurate representation of the effective surface gravity. Specifically, it accounts for the centrifugal force resulting from the stellar rotation, which directly impacts the effective gravity and is crucial to accurately deriving the stellar mass. Despite this correction, we caution that uncertainties may still be underestimated.
The wind parameters of the Be star were constrained on the basis of diagnostic lines in the UV, in particular the \ion{Si}{iv\,$\lambda\lambda1394, 1403$} doublet. The absence of P Cygni-like line profiles indicates moderate wind strength, and we found a best-fit mass loss rate $\log \dot{M} = -8.5 \pm 0.2\, M_\odot \mathrm{yr}^{-1}$ at fixed terminal wind velocity of $\varv_{\infty}=800$\,km\,s$^{-1}$. We note that non-spherical wind structures, expected in Be stars and not captured by 1D models, may affect the detailed line shape and velocity centroid \citep[e.g.][]{Puls+2008}.

Building on the analysis of the fundamental parameters, we further estimated the CNO surface abundances in the Be star. We found that the observed optical and UV absorption lines for these elements did not match models assuming solar abundances. Specifically, the lines for carbon and oxygen appeared weaker, while those for nitrogen appeared stronger. To better fit the observed absorption lines, we adjusted the CNO abundances in our models (see Table~\ref{tab:stellar_abundances}). The observed spectra are better reproduced by a model with a factor of three higher nitrogen abundance and factors of three and two lower abundances for carbon and oxygen, respectively. The CNO-processed material on the surface of the Be star could be the result of its rapid rotation, or past mass transfer from the companion star \citep[e.g.][though see also \citealt{Hunter+2008}]{Maeder_Meynet2000,Langer+2008}.

In OB stars, instabilities in line‑driven winds can produce embedded shocks that emit soft X‑rays \citep[e.g.][]{Zhekov_Palla2007}.
Our PoWR models include an X-ray luminosity at the observed level of $L_\mathrm{X} \sim 10^{30}$ erg\,s$^{-1}$ \citep{Ribo+2017}, which has little impact on the derived Be-star stellar parameters but slightly improves the fit to some UV wind features (see Sect.~\ref{sec:UV_signs_of_companion_wind}).
In PoWR, X-rays are treated as free-free emissivity added into the radiative transfer calculations \citep{Baum+1992}. The X-rays are parameterised by temperature $T_\text{X}$, which we set to $1\,$MK in this work \citep[$\sim 0.1$\,keV, typical for OB star winds,][]{Zhekov_Palla2007}, and a filling factor adjusted to match the observed $L_\text{X}$.

\section{Signs of the companion in the UV}
\label{sec:UV}

\begin{figure*}[ht!]
    \centering
    \includegraphics[width=\textwidth]{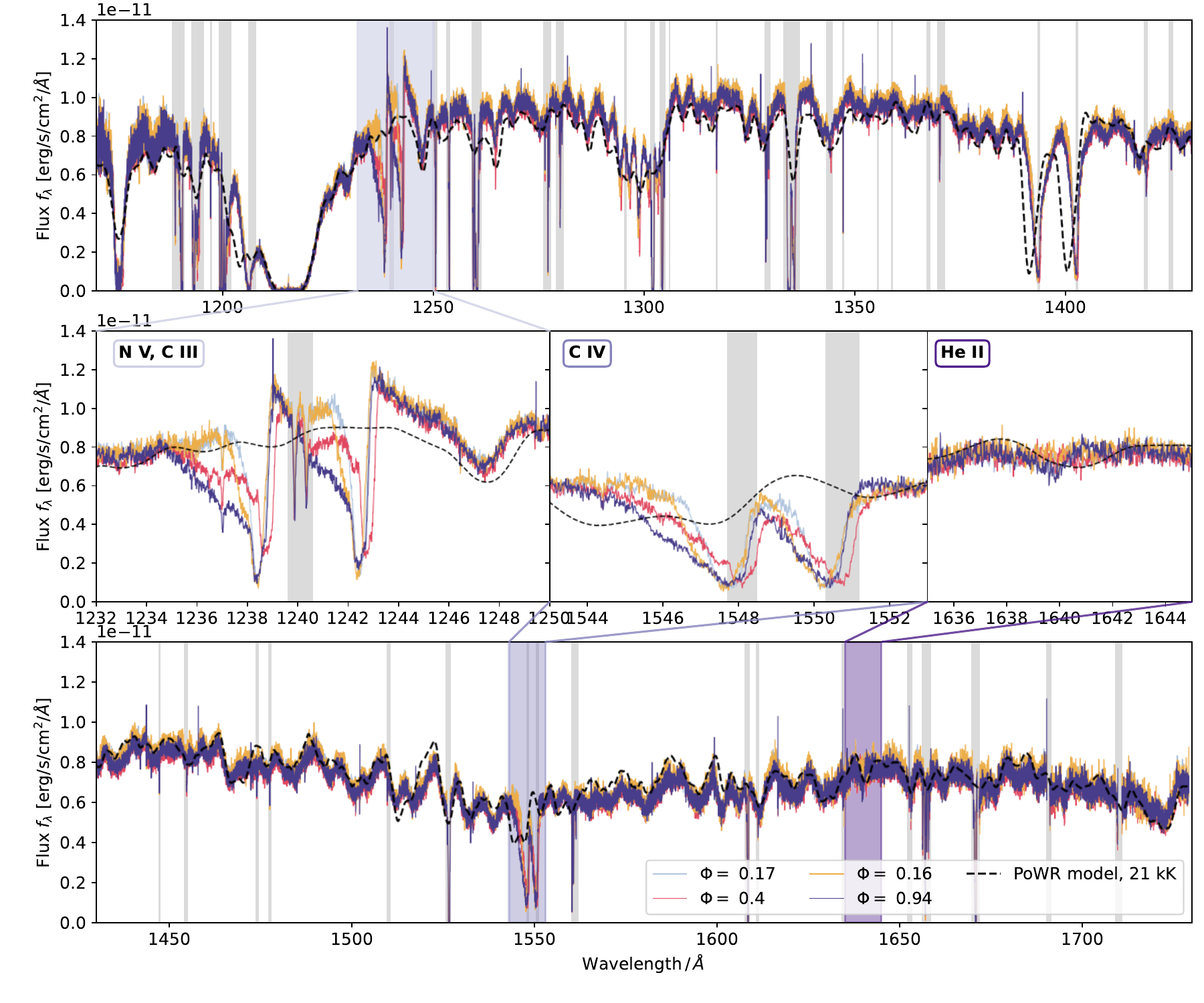}
    \caption{HST/STIS UV spectra of MWC~656 obtained on four observing dates. A PoWR stellar atmosphere model with $T_* = 21$\,kK, $\log g_* = 3.4$, and $v\sin i = 300$\,km\,s$^{-1}$ is shown in black for comparison. The top and bottom panels show the full spectral range, split into blue and red halves, while the middle row displays zoom-ins on key diagnostic lines: the \ion{N}{v\,$\lambda\lambda$1238, 1241} doublet, the \ion{C}{iv\,$\lambda\lambda$1548, 1551} doublet, and the \ion{He}{ii\,$\lambda$1640} line. Shaded grey regions mark known ISM absorption features. The zoom-in panels reveal prominent high-ionisation features, not present in the Be star model, suggesting they originate from a hotter source within the system.} 
    \label{fig:spec_uv_zoom}
\end{figure*} 

The HST/STIS FUV spectra, obtained at four orbital phases (two near opposite quadratures at $\Phi = 0.4,\,0.94$, and two near conjunction, when the Be star was in front of the companion along the line of sight at $\Phi = 0.16,\,0.17$; see Fig.~\ref{fig:orbit_fit}), are displayed in Fig.~\ref{fig:spec_uv_zoom}. The Be star model reproduces most of the FUV spectrum well, confirming that the observed flux is dominated by the Be star. However, several features are not reproduced by the model and instead suggest the presence of a hotter component in the system. We discuss these features in the following.

\subsection{\ion{He}{ii} emission}
\label{sec:UV_signs_of_companion_Heii}
The \ion{He}{ii\,$\lambda1640$} line, shown in the rightmost zoom panel of Fig.~\ref{fig:spec_uv_zoom}, displays clear phase-dependent variability (see also Fig.~\ref{fig:disentangling_UV}). Mid- to late-type B stars generally do not exhibit \ion{He}{ii} lines, and therefore the variability seen here is unlikely to originate from the Be star. Despite the limited number of available epochs, we detect an emission component that shifts in anti-phase with the Be star, consistent with the companion (see Appendix~\ref{app:disentangling}). This behaviour mirrors our findings in the optical, where the \ion{He}{ii\,$\lambda4686$} line also appears in emission, tracking the companion.

\subsection{High-ionisation wind lines}
\label{sec:UV_signs_of_companion_wind}

\begin{figure*}[t]
    \centering
    \includegraphics[width=\textwidth]{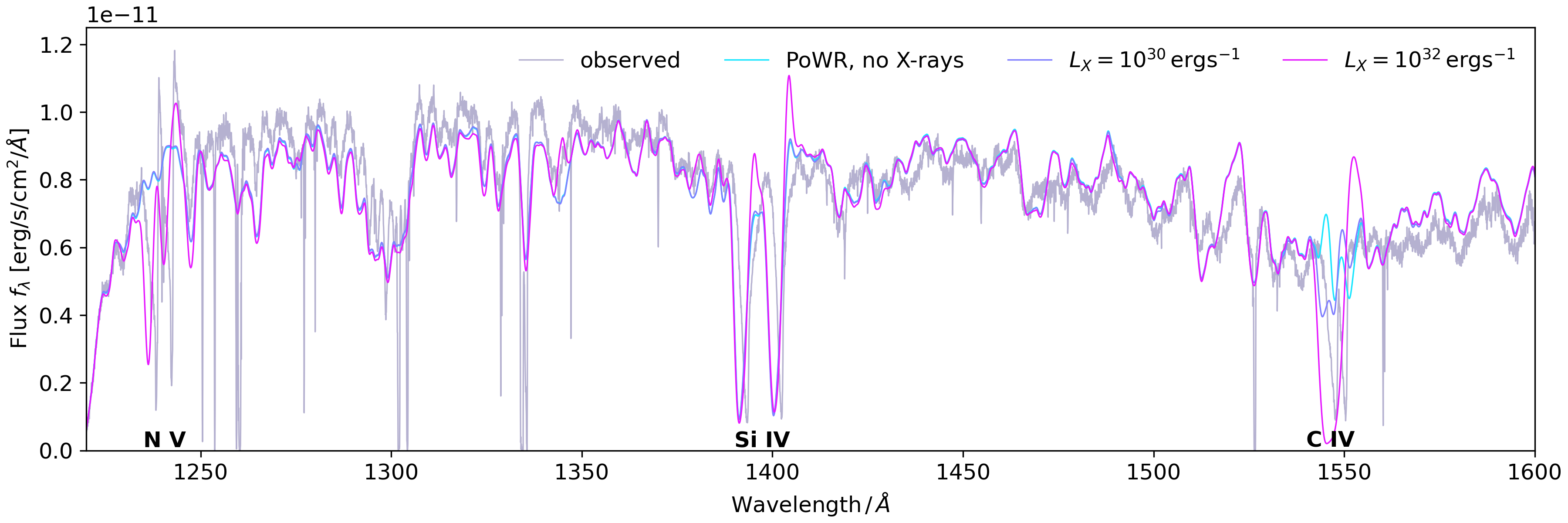}
    \caption{Observed spectrum of MWC~656 (grey) compared to synthetic Be star spectra computed for different X-ray luminosities, from no X-rays up to $10^{32}$\,erg\,s$^{-1}$. Our fiducial model assumes $L_\mathrm{X} = 10^{30}$\,erg\,s$^{-1}$, consistent with observations \citep{Ribo+2017}. The HST spectrum has been re-binned to 3 pixels for clarity. Only the $10^{32}$\,erg\,s$^{-1}$ model produces \ion{N}{v} emission comparable to the data, but it simultaneously generates \ion{Si}{iv} and \ion{C}{iv} features that are not observed.} 
    \label{fig:model_xrays}
\end{figure*} 

Beyond \ion{He}{ii}, the spectra show several high-ionisation species, shown in Fig.~\ref{fig:spec_uv_zoom}. 
The \ion{N}{v\,$\lambda\lambda$1239, 1243} doublet exhibits prominent P Cygni profiles, especially near quadrature, with blueshifted absorption and redshifted emission components. This suggests a hot, radiatively driven wind. Although the \ion{Si}{iv\,$\lambda\lambda$1393, 1402} and \ion{C}{iv\,$\lambda\lambda$1548, 1551}  lines do not show clear emission components, they exhibit sloped blueshifted absorption troughs indicative of wind outflows, especially at $\Phi=0.4$ and $0.94$. These wind features coincide with an interstellar absorption component.

The \ion{Si}{iv} and \ion{C}{iv} lines can arise from super-ionisation in B-star winds, typically interpreted as the result of shock-generated X-rays that raise ionisation levels above those expected for the stellar temperature \citep[e.g.][]{Puls+2008}. Such effects are common in Be stars and explain their stronger UV wind lines compared to non-emission B stars \citep{Grady+1987, Prinja1989}. Indeed, incorporating the observed level of X-rays into our 21\,kK Be-star model reproduces the absorption components of the \ion{Si}{iv} and \ion{C}{iv} resonance lines. However, even accounting for super-ionisation, \ion{N}{v} P-Cygni emission is only observed in the most luminous Be stars earlier than B1 \citep{Kogure_Hirata1982,Rivinius+2013}. As discussed in Sect.~\ref{sec:discussion_comparison}, no comparable \ion{N}{v} emission is seen in other Be stars of similar type to MWC~656. Producing such features in MWC~656 therefore requires either a substantially hotter ionising source or an unusually high X-ray luminosity.

Figure~\ref{fig:model_xrays} illustrates how the strength of \ion{N}{v} in the Be star model responds to increasing X-ray flux. At MWC 656's observed X-ray luminosity ($L_\mathrm{X} \sim 10^{30}$\,erg\,s$^{-1}$), no detectable \ion{N}{v} emission is produced. Only at X-ray luminosities exceeding $10^{32}$\,erg\,s$^{-1}$ -- two orders of magnitude above the observed level -- do the synthetic spectra show \ion{N}{v} features comparable to observations.
In addition to being incompatible with the quiescent state of the system, these high X-ray flux models also produce strong \ion{C}{iv} and \ion{Si}{iv} emission lines that are not observed in MWC~656, creating additional inconsistencies with the data. Here, we kept the Be star mass-loss rate fixed to the value derived in Sect.~\ref{sec:BeStar}; increasing $\dot{M}$ could in principle enhance \ion{N}{v}, but would also produce \ion{Si}{iv} and \ion{C}{iv} emission components, which are not seen. Therefore, X-ray heating of the Be star wind seems an unlikely explanation for the observed UV features. Instead, the observed \ion{N}{v} doublet and the contributions to \ion{He}{ii} and \ion{C}{iv}, are more naturally explained by the presence of a hot, luminous companion and its associated wind.

\subsection{Doppler shifts}
\label{sec:UV_signs_of_companion_doppler}
Using the orbital solution presented in Sect.~\ref{sec:binary_orbit}, we predicted Doppler shifts for both components across the observations. For the Be star, the expected maximum RV amplitude between observations is approximately 26\,km\,s$^{-1}$.
To estimate RV shifts from the observed UV spectra, we performed an iterative cross-correlation of each epoch against the $\Phi=0.16$ spectrum, focussing on wavelength regions free from strong interstellar absorption and wind lines. This yields relative RVs of $-12 \pm 5$, $-1 \pm 5$, and $-1 \pm 5$\,km\,s$^{-1}$ for $\Phi=0.94, 0.4$, and $0.17$, respectively. The observed shift at $\Phi=0.94$ is qualitatively consistent with the expected lower RV at this phase, although the measured amplitude is smaller than predicted. However, we note that, given the moderate S/N, broad spectral features, and relatively low RV amplitude of the Be star, precise measurements are challenging.
The companion is predicted to have an RV amplitude of about 140\,km\,s$^{-1}$ between observations. However, the FUV spectrum is dominated by the Be star, making it challenging to directly measure the companion’s RVs. Nevertheless, we recover He~\textsc{ii} emission from the companion through spectral disentangling using the predicted orbital velocities (Fig.~\ref{fig:disentangling_UV}), providing indirect support for the expected motion. A puzzling feature of the UV observations is the apparent Doppler shift seen in the high-ionisation wind lines, which follows neither of the binary components and is difficult to explain in any binary companion scenario. We address this aspect in more detail in Sect.~\ref{sec:discussion_caveats}.

\subsection{Phase-dependent spectral variability}
\label{sec:UV_spectral_variability}
The four epochs of HST/STIS FUV spectroscopy reveal phase-dependent variability in the high-ionisation UV wind lines. Resonance line profiles appear to strengthen near quadrature, with, for example, the \ion{C}{iv} doublet showing more prominent blueshifted absorption troughs during these phases, while features weaken near conjunction ($\Phi = 0.16$--0.17). Variability in the line profiles of resonance lines such as \ion{Si}{iv} and \ion{C}{iv} is common in Be stars \citep[e.g.][]{Grady+1987}. However, the observed changes in the UV line profile may also reflect the interaction between the companion and the Be disc. In binaries, hydrodynamic simulations show that the companion can perturb the Be disc, exciting spiral density waves that produce phase-dependent variations in emission and absorption \citep{Panoglou+2016, Rubio+2025}. The phase-dependence could also indicate ionisation of the Be star wind by the hot companion, in which case the absorption would originate on the far side of the system during conjunction, or the wind of the hot companion itself, whose signatures diminish when it is partially obscured behind the Be star. However, we note that with only four UV epochs, some apparent phase-dependence could be coincidental.

\section{Composite model with a stripped-star companion}
\label{sec:model}

The UV data revealed spectral features, such as high ionisation wind lines in \ion{N}{v} and \ion{He}{ii} emission, absent in the Be star spectrum, which point towards the presence of a hotter luminous companion in the system (Sect.~\ref{sec:UV}). Motivated by these findings, we modelled the binary as a composite of two luminous stars, featuring the Be star and a stripped helium star companion. We performed a combined spectroscopic and photometric analysis, with a setup similar to that in Sect.~\ref{sec:BeStar}, with the companion star parameters optimised to reproduce the observed UV wind lines.

\begin{figure*}[ht]
    \centering
    \includegraphics[width=0.8\textwidth]{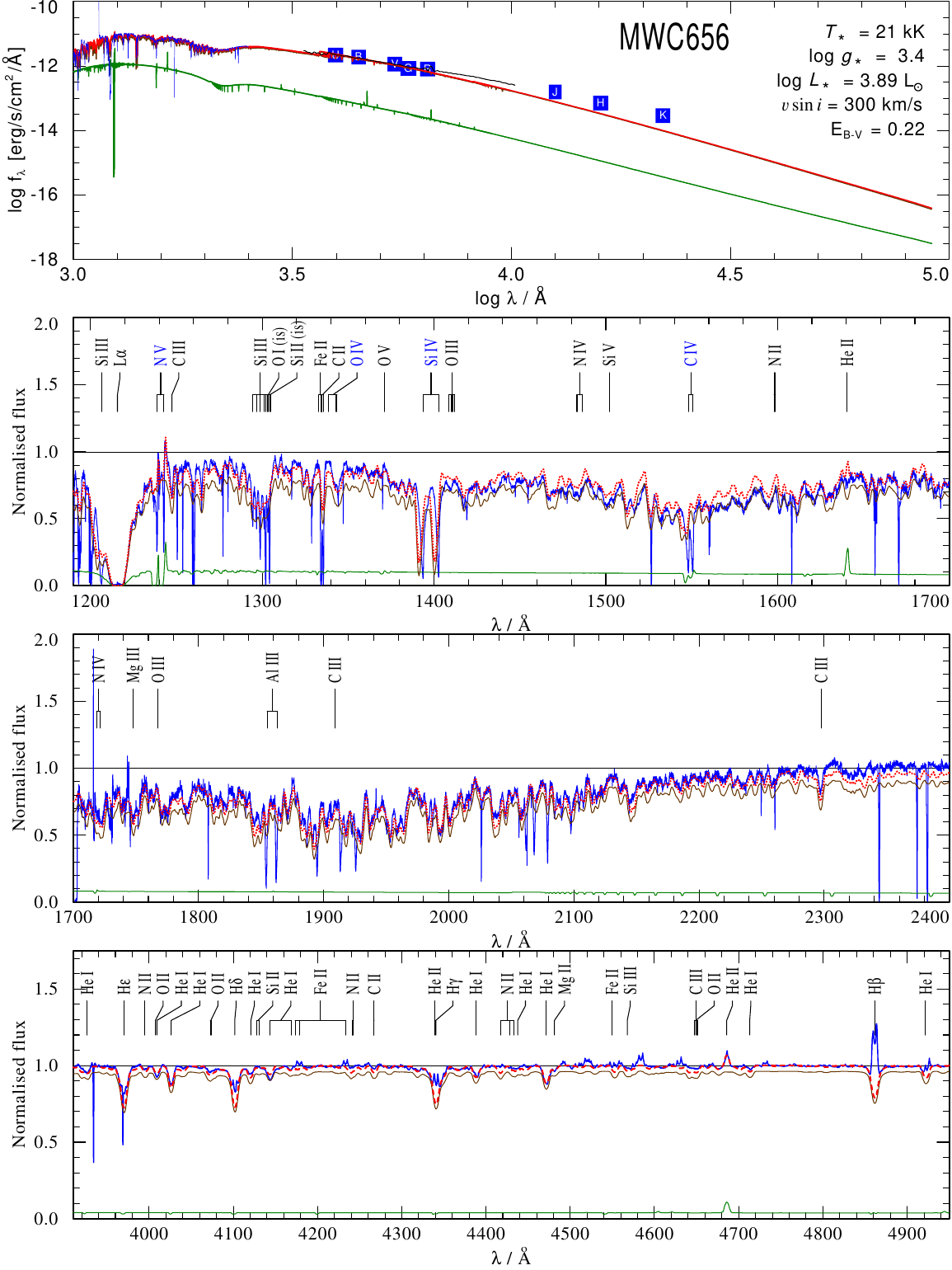}
    \caption{Observed spectra and photometry of MWC~656 (blue) compared to the model SED and synthetic spectra. The composite model (red) is a weighted sum of the rapidly rotating Be star (brown) and the hot stripped companion star (green). The \textit{Gaia} XP spectrum is overplotted in the top panel (black). The most notable features of the stripped companion are its contributions to the N~\textsc{v} and He~\textsc{ii} lines. The infrared excess seen in the top panel originates in the circumstellar disc of the Be star. The narrow absorption lines in the UV and optical spectra result from interstellar absorption.} 
    \label{fig:composite_spectral_fit}
\end{figure*} 

\begin{table}[t]
        \caption{Parameters derived for MWC~656 from composite spectroscopic analysis, assuming the observed UV features originate from a hot luminous companion.}
    \label{tab:stellar_parameters}
        \centering
        \renewcommand{\arraystretch}{1.6}
        \begin{tabular}{lcc}
                \hline
                \hline
                \vspace{0.1cm}
                &       Be star &       Stripped star \\
                \hline
                $T_{\ast}$ (kK) & $21 \pm 1$ & $85^{+10}_{-5}$ \\
                $T_{2/3}$ (kK) & $20.8 \pm 1$ & $82.7^{+10}_{-5}$ \\
                $\log g_\ast$ (cm\,s$^{-2}$) & $3.4 \pm 0.1$ & $5.3 \pm 0.3$ \\
                $\log g_{2/3}$ (cm\,s$^{-2}$) & $3.38 \pm 0.1$ & $5.26 \pm 0.3$ \\
                $\log g_\mathrm{true}$\tablefootmark{\dag} (cm\,s$^{-2}$) & $3.65 \pm 0.1$ & $5.3 \pm 0.3$ \\
                $\log L$ ($L_\odot$) & $3.9 \pm 0.1$ & $4.0 \pm 0.2$ \\
                $R_\ast$ ($R_\odot$) & $6.8 \pm 1.0$ & $0.51 ^{+0.16}_{-0.14}$ \\
                $\log \dot{M}$ ($M_\odot \mathrm{yr}^{-1}$) & $-8.5 \pm 0.2$ & $-7.5 \pm 0.2$ \\
                $\varv_{\infty}$ (km\,s$^{-1}$) & 800 (fixed) & $400 \pm 100$ \\
                $\varv \sin i$ (km\,s$^{-1}$) & $300 \pm 50$ & 10 (fixed) \\
                $\varv_{\mathrm{mac}}$ (km\,s$^{-1}$) & $150 \pm 30$ & 10 (fixed) \\
                $X_{\rm H}$ (mass fr.) & 0.737 & 0.2 (fixed)\\
                $X_{\rm He}$ (mass fr.) & 0.26 & 0.79 (fixed)\\
                $M_\mathrm{spec}$ ($M_\odot$) & $7.4 \pm 2.7$ & $1.9  ^{+1.7}_{-1.6}$ \\
                $\log\,Q_{\mathrm H}$ (s$^{-1}$) & 46 & 48 \\
                $\log\,Q_{\mathrm {He\,\textsc{ii}}}$ (s$^{-1}$) & - & 45 \\
        \hline
        $E_{\mathrm{B-V}}$ (mag) & \multicolumn{2}{c}{$0.22 \pm 0.02$} \\
                $R_V$ & \multicolumn{2}{c}{$2.45 \pm 0.05$} \\
                \hline
        \end{tabular}
        \tablefoot{$\log\,Q_{\mathrm{H}}$, $\log\,Q_{\mathrm {He\,\textsc{ii}}}$ denote the inferred emission rates of H- and He~\textsc{ii}-ionising photons. \tablefoottext{\dag} {$\log (g_\ast +(\varv \sin i)^{2} /R_\ast)$}}
\end{table}

For the Be star, we fixed the fundamental parameters ($T_\ast$, $\log g_\ast$) to the values obtained in Sect.~\ref{sec:BeStar}, but allowed the luminosity, and hence the radius, to vary. The companion was also modelled using PoWR atmospheres, assuming a hydrogen-depleted ($X_\mathrm{H}=0.2$) and helium-enriched ($X_\mathrm{He}=0.79$) composition, typical of known stripped stars \citep[e.g.][]{Ramachandran+2023, Ramachandran+2024, Goetberg+2023, Mueller-Horn+2025}. The stripped nature of the companion is motivated by its inferred properties -- hotter but less massive than the Be star -- which are consistent with a post-mass-transfer binary \citep[e.g.][]{Wang+2021,Drout+2023}.
The free parameters were its effective temperature, surface gravity, luminosity, mass-loss rate, and terminal wind velocity. Because no clear photospheric absorption lines are visible, we fixed its rotational and macroturbulent velocities to 10\,km\,s$^{-1}$.

The companion's effective temperature is primarily constrained by the observed UV emission lines. The presence of \ion{N}{v} requires a minimum temperature of $\approx40\,$kK. However, reproducing the observed \ion{N}{v} emission without accompanying strong \ion{C}{iv} or \ion{Si}{iv} emission requires significantly higher temperatures, $T_{\ast,2}\gtrsim80$\,kK, consistent with trends seen in Wolf-Rayet stars \citep[e.g.][]{Willis+1986,Hamann+2004,Hamann+2006,Sander+2025}.

We explored companion temperatures from 65 to 95\,kK, with corresponding $\log g_{\ast,2}$ scaled between 4.6 and 5.6, to be consistent with the helium zero-age main sequence (ZAMS). Models with lower temperatures produce stronger \ion{C}{iv} and \ion{Si}{iv} lines but weaker \ion{N}{v} features. In contrast, models that are too hot have a steep SED, which reduces their contribution to the optical spectrum. By balancing the strength and shape of the \ion{N}{v} P Cygni profile with the contribution to optical lines such as \ion{He}{ii\,$\lambda$4686}, our models suggest a best-fit temperature of $T_{\ast,2} = 85^{+10}_{-5}$\,kK. The sensitivity of the fit to the assumed temperature is illustrated in Appendix~\ref{app:stripped_star}, where we compare cooler and hotter companion models to the observations. 

The resulting best-fit composite model is shown in Fig.~\ref{fig:composite_spectral_fit}, with key parameters summarised in Table~\ref{tab:stellar_parameters}. We derive $E(B-V) = 0.22\pm0.02$\,mag and $R_V = 2.45\pm0.05$. We note that the corresponding $V$-band extinction of $A_V = 0.54\pm 0.04$\,mag agrees well with 3D dust map predictions \citep[e.g. $A_V \approx 0.53\pm 0.07$ from Bayestar19;][]{Green2018,Green+2019}, suggesting negligible system-intrinsic extinction. The Be star has an inferred luminosity of $\log L_{\mathrm{Be}}/\mathrm{L}_\odot = 3.9\pm0.1$ and a stellar radius of $6.8 \pm 1.0 \, R_\odot$, the latter determined using the Stefan-Boltzmann law. As expected, attributing part of the observed flux to the stripped companion results in slightly lower inferred luminosity and radius compared to the single-star model (Sect.~\ref{sec:BeStar}), but the differences are minimal. 
This is because the companion's contribution to the total flux is small, and the Be star dominates both observed UV and optical light.

Our model suggests that the stripped companion has a luminosity of $\log L_{2}/\mathrm{L}_\odot \approx 4.0$, comparable to that of the Be star, and a radius of $R_{\ast,2} \approx 0.5$\,R$_\odot$. The companion luminosity is constrained by the composite SED and the absence of detectable photospheric absorption in the spectrum. The strength and shape of the \ion{N}{v} P Cygni profile further constrain the companion’s wind parameters, yielding a mass-loss rate of $\log \dot{M}_\mathrm{2} \approx -7.5 $\,M$_\odot$\,yr$^{-1}$, about an order of magnitude higher than that of the Be star. A terminal wind velocity of $v_\infty \simeq 400~\mathrm{km\,s^{-1}}$ is estimated from the blue absorption edge of the \ion{N}{v} P-Cygni profile. This diagnostic may provide a lower limit to the true $\varv_\infty$, since the maximum velocity extent seen in absorption varies between epochs, plausibly due to interaction of the wind with the Be-star companion. We note that the derived $\varv_\infty$ is very low compared to the empirical correlations observed for O-type stars between $\varv_\infty$, $T_\mathrm{eff}$, and escape velocity \citep[e.g.][]{Hawcroft2024}. However, in the absence of prior measurements of terminal wind velocities for hot, intermediate-mass stripped stars, it remains unclear whether these objects follow the same correlations established for OB stars. It should be noted that the present object occupies a largely unexplored regime of hot ($\sim$$80$~kK) stripped stars. In this regime, the atmosphere structure and the resulting winds are expected to be strongly affected by radiatively driven turbulence associated with the hot iron opacity bump \citep[e.g.][]{Moens2022,Debnath2024}. A quantitative assessment of this effect is beyond current 1D atmosphere modelling capabilities and thus beyond the scope of this paper. The companion contributes roughly 10--15\% of the flux in the UV and less than 5\% in the optical.
We note that these companion parameters were derived under the assumption that the system contains a hot luminous star with a wind responsible for the \ion{N}{v} and \ion{He}{ii} emission features. This model does not account for potential contributions from circumstellar discs or interactions between the stellar winds of the two components, which could affect the inferred parameters (Sect.~\ref{sec:discussion_caveats}).

\section{Nature of the companion}
\label{sec:companion}

Having established the orbital parameters, Be star properties, and spectral modelling in the preceding sections, we now bring together all available evidence to assess the nature of the unseen companion. We first combine dynamical, spectroscopic, and evolutionary constraints to narrow down its plausible mass range, and then evaluate whether compact objects of such mass could account for the observed spectral features.

\subsection{Companion mass}
\label{sec:companion_mass}

\begin{figure}[t]
    \centering
        \includegraphics[width=\columnwidth]{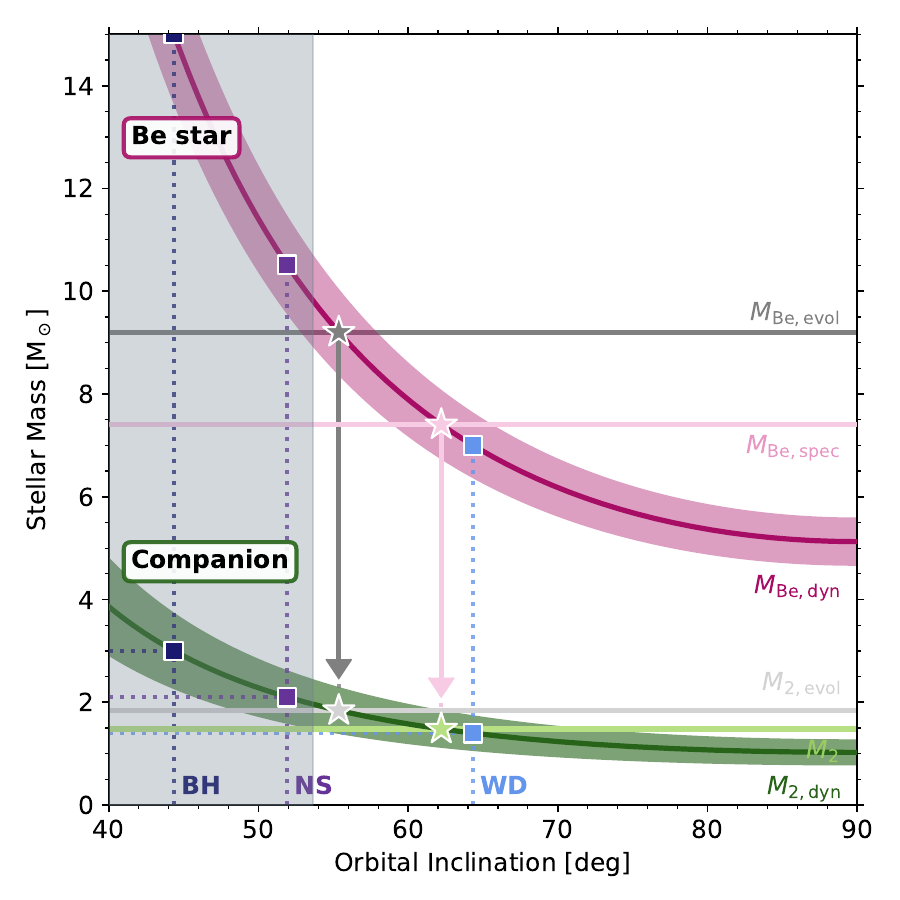}
   \caption{Stellar mass constraints for the Be star in MWC~656 and its companion. Dynamical mass constraints are shown as a function of orbital inclination for the Be star (magenta curve) and its companion (green curve) with shaded regions indicating $1\sigma$ uncertainties. Horizontal lines mark independent mass estimates: the Be star’s evolutionary mass $M_{\mathrm{Be,evol}} = 9.20^{+0.61}_{-0.55}\,\mathrm{M}_\odot$ (grey; upper limit), and spectroscopic mass estimate; $M_{\mathrm{Be,spec}} = 7.4 \pm 2.7\,\mathrm{M}_\odot$ (solid pink line). The companion mass estimates, $M_{2} = 1.48^{+0.55}_{-0.46}\,\mathrm{M}_\odot$ and $M_{2,\mathrm{evol}} = 1.84 \pm 0.13 \,\mathrm{M}_\odot$, inferred from combining the dynamical mass ratio with spectroscopic and evolutionary estimates of the Be star mass, are shown in light green and light grey, respectively. The grey hatched region indicates inclinations where the dynamical Be star mass would exceed its evolutionary upper limit by more than $1\sigma$, marking improbable configurations. Dotted lines denote the minimum inclinations required for different companion types: 1.4\,M$_\odot$ for a white dwarf (light blue), 2.1\,M$_\odot$ for a neutron star (purple), and 3.0\,M$_\odot$ for a BH (navy). A BH companion would require $i \lesssim 45^\circ$ and Be star masses $>15\,\mathrm{M}_\odot$, which is strongly disfavoured. The intersection of constraints confines the likely companion mass to $1.0$--$1.9\,\mathrm{M}_\odot$.}
    \label{fig:be_star_stellar_tracks}
\end{figure}

The mass of the companion in MWC~656 can be constrained by combining the dynamical mass ratio with spectroscopic and evolutionary estimates of the mass of the Be star.

From the orbital solution, we obtained a dynamical mass ratio of $q_\mathrm{dyn} = 5.25$ and minimum masses of
$$M_\mathrm{Be} \sin^3 i = 5.13\pm0.47\,\mathrm{M}_\odot, \ M_\mathrm{2} \sin^3 i = 1.03\pm0.26\,\mathrm{M}_\odot\,,$$
which is illustrated in Fig.~\ref{fig:be_star_stellar_tracks}.
The Be star mass inferred from our composite spectroscopic model is $M_\mathrm{Be,spec} = 7.4 \pm 2.7\,\mathrm{M}_\odot$ (The mass obtained from the single-star analysis was $\sim$8 $M_\odot$). Combining this with the mass function implies an inclination of $62^{+28}_{-10}$ degrees and a companion mass of $M_\mathrm{2} = 1.48^{+0.55}_{-0.46}\,\mathrm{M}_\odot$.

An evolutionary mass was also derived by comparing the effective temperature and luminosity of the Be star with 
simulated evolutionary tracks. We used the \textsc{BONNSAI}\footnote{The \textsc{BONNSAI} web-service is available at \url{www.astro.uni-bonn.de/stars/bonnsai}} code \citep{Schneider+2014}, which implements a Bayesian statistical method to determine probability distributions of fundamental stellar parameters. We used the massive star evolutionary model grids by \cite{Brott+2011} and flat prior distributions in stellar mass and age. This yields $M_\mathrm{Be,evol} = 9.20^{+0.61}_{-0.55}\,\mathrm{M}_\odot$, which corresponds to a companion mass of $M_\mathrm{2,evol} \approx 1.84 \pm 0.13\,\mathrm{M}_\odot$, consistent with the spectroscopic mass estimate. However, this value may represent an upper limit, as past mass accretion could make the Be star temporarily appear over-luminous relative to its true mass \citep[e.g.][]{Kippenhahn_Meyer-Hofmeister1977,Lau+2024}. 

Together, dynamical, spectroscopic, and evolutionary constraints consistently place the companion mass between 1.0 and 1.9\,M$_\odot$.
Two key conclusions follow:
First, a BH companion is strongly disfavoured. The inferred mass lies well below the canonical lower mass threshold for BHs ($\gtrsim$3\,M$_\odot$). For comparison, \citetalias{Casares+2014} proposed a BH mass of 3.8--6.9\,M$_\odot$. Even accommodating the lowest of these values would require a Be star mass of $\sim$19\,M$_\odot$, which is ruled out at $>4\sigma$ by both spectroscopic and evolutionary estimates.
Second, a white dwarf companion is also unlikely. Even the minimum allowed dynamical companion mass at $i = 90^\circ$ ($M_\mathrm{2} \sin^3 i = 1.03\pm0.26$\,M$_\odot$) implies a very massive white dwarf, and the most likely companion mass of $1.48^{+0.55}_{-0.46}$\,M$_\odot$ exceeds the Chandrasekhar limit ($\sim$1.4\,M$_\odot$), placing most of the plausible mass range above this threshold.

In summary, the combined mass constraints effectively rule out a BH and make a white dwarf companion improbable, favouring instead a neutron star or an intermediate-mass stripped star.

\subsection{Could the companion still be an accreting compact object?}
\label{sec:companion_compact_object}

While the inferred companion mass of $M_\mathrm{2} =  1.48^{+0.55}_{-0.46}\,\mathrm{M}_\odot$ disfavours a white dwarf and excludes a BH, we briefly examine whether an accreting compact object, either a neutron star or, less likely, a massive white dwarf, could reproduce the observed UV wind features.

If the companion was a dark, compact object, the UV lines would need to originate in an accretion disc and its outflows. A neutron star is the most viable option in terms of mass. Accretion disc outflows can in fact produce strong emission in \ion{N}{v} and \ion{He}{ii} with P Cygni-like profiles \citep[e.g.][]{CastroSegura+2022,Fijma+2023}. However, such systems are typically bright X-ray sources, whereas MWC~656 is not -- its X-ray luminosity is over four orders of magnitude below that of Be+neutron star binaries with similar orbits \citep{Brown+2018}. It is therefore difficult to reconcile the observed UV lines with an accretion disc powerful enough to drive them.

A white dwarf accretor provides another possible explanation. X-rays from accretion could photo-ionise the Be-star wind and produce \ion{N}{v} emission. This scenario has been proposed to explain the so-called $\gamma$~Cas stars, Be stars with anomalously high X-ray luminosities that may host accreting white dwarfs \citep[e.g.][]{Smith+2016,Gies+2023}. Although MWC~656 is significantly less X-ray luminous than typical $\gamma$~Cas stars, we tested this scenario by incorporating increasing levels of X-ray heating in the Be-star model (Sect.~\ref{sec:UV_signs_of_companion_wind}). Reproducing the observed \ion{N}{v} strength requires X-ray luminosities about two orders of magnitude higher than observed in MWC~656, and the models simultaneously predict strong \ion{C}{iv} and \ion{Si}{iv} emission that is not seen in the data.

In conclusion, neither a neutron star nor a white dwarf accretor can explain the observed UV wind features given the low X-ray luminosity and spectral constraints of the system. The stripped-star companion thus remains the most consistent interpretation. Although unlikely to be the origin of the UV wind signatures, some degree of mass transfer from the Be disc onto the stripped companion remains possible. The potential presence and implications of a circumsecondary accretion disc are discussed in Appendix~\ref{app:circumsecondary_disc}.

\section{Discussion}
\label{sec:discussion}

\subsection{Comparison of MWC~656 to known stripped stars}
\label{sec:discussion_comparison}

\begin{figure*}[ht!]
    \centering
    \begin{minipage}[t]{\textwidth}
        \centering
        \includegraphics[width=\textwidth]{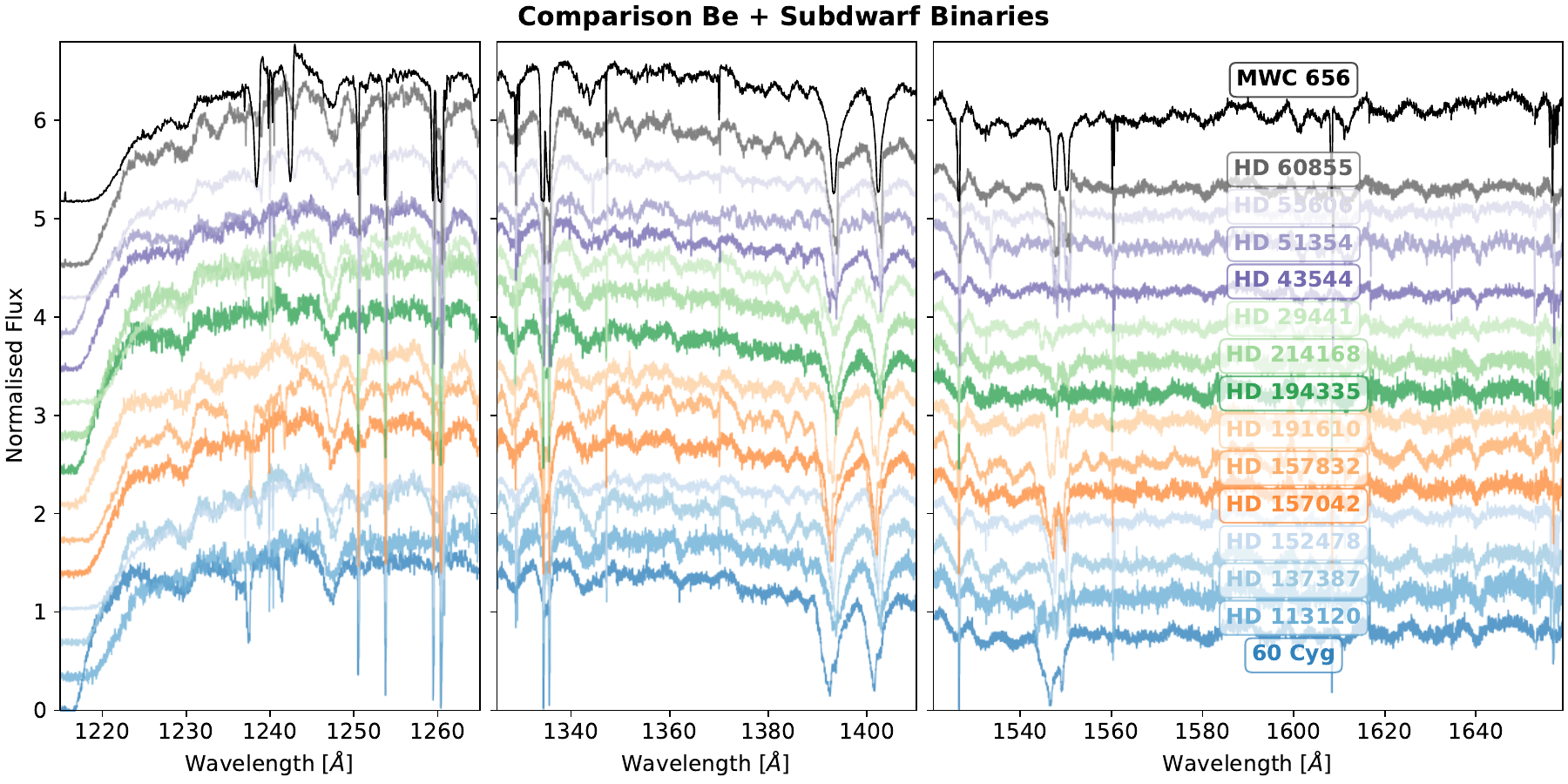}
    \end{minipage}
    \begin{minipage}[t]{\textwidth}
        \centering
        \includegraphics[width=\textwidth]{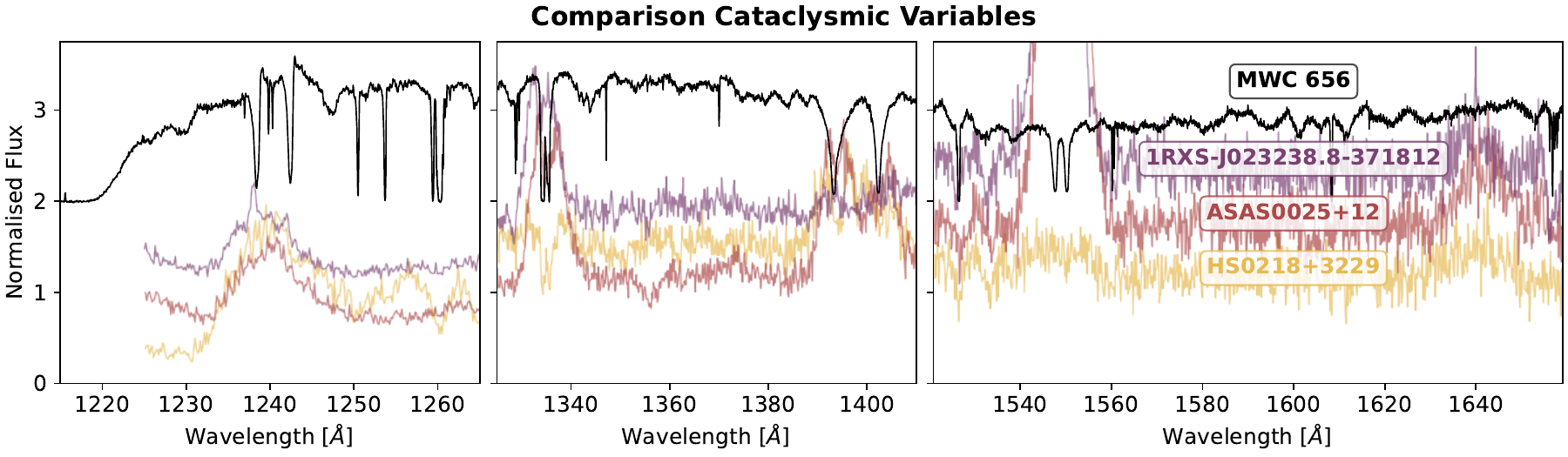}
    \end{minipage}
    \caption{Comparison of the HST/STIS UV spectrum of MWC~656 (black) with reference spectra from two classes of stars: known Be+subdwarf systems \citep[top;][]{Wang+2021}
    and cataclysmic variables \citep[bottom;][]{Pala+2022,Toloza+2023}. The Be+subdwarf and cataclysmic variable spectra have been re-binned using bin sizes of three and two, respectively, to improve S/N. All spectra have been normalised by their median flux, and vertical offsets have been applied for readability.}
    \label{fig:stripped_star_comparison_spectra}
\end{figure*}

\begin{figure}[t]
    \centering
    \includegraphics[width=\columnwidth]{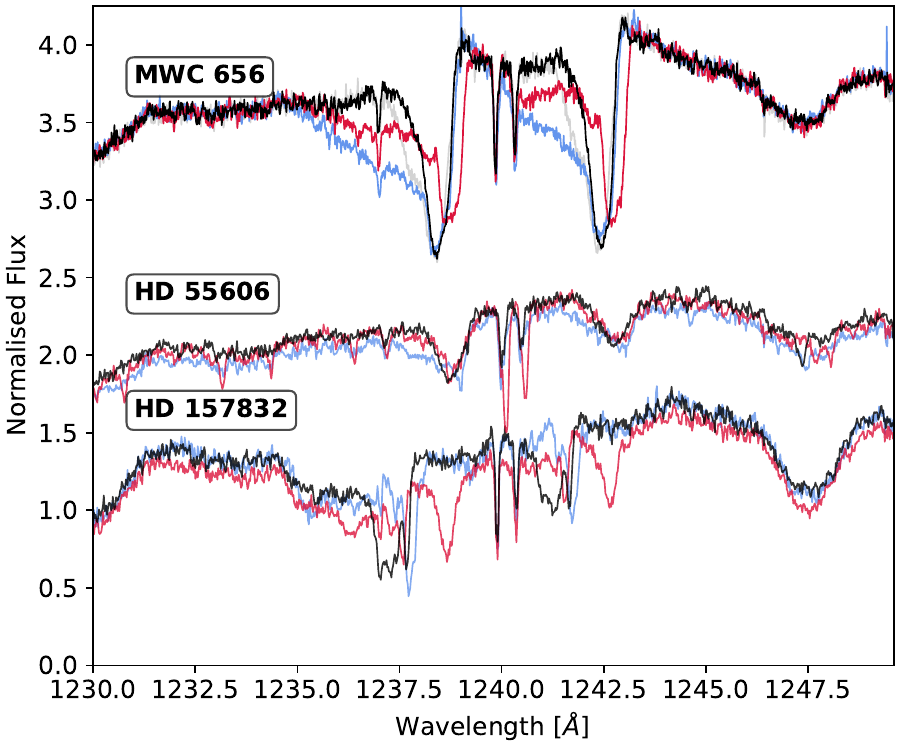}
    \caption{Comparison of HST/STIS UV spectra of MWC 656 (top) with those of two candidate Be+subdwarf binaries, HD~55606 and HD~157832 \citep{Wang+2021}. Different colours represent different observational epochs. All spectra are normalised by their median flux, and vertical offsets have been applied for clarity. Narrow lines near $1240\,\AA$ originate from ISM absorption. Like MWC~656, both comparison systems exhibit line profile variability in N~\textsc{v} and possible signatures of stellar winds.} 
    \label{fig:stripped_star_comparison_sdO}
\end{figure} 

\begin{figure}[t]
    \centering
    \includegraphics[width=\columnwidth]{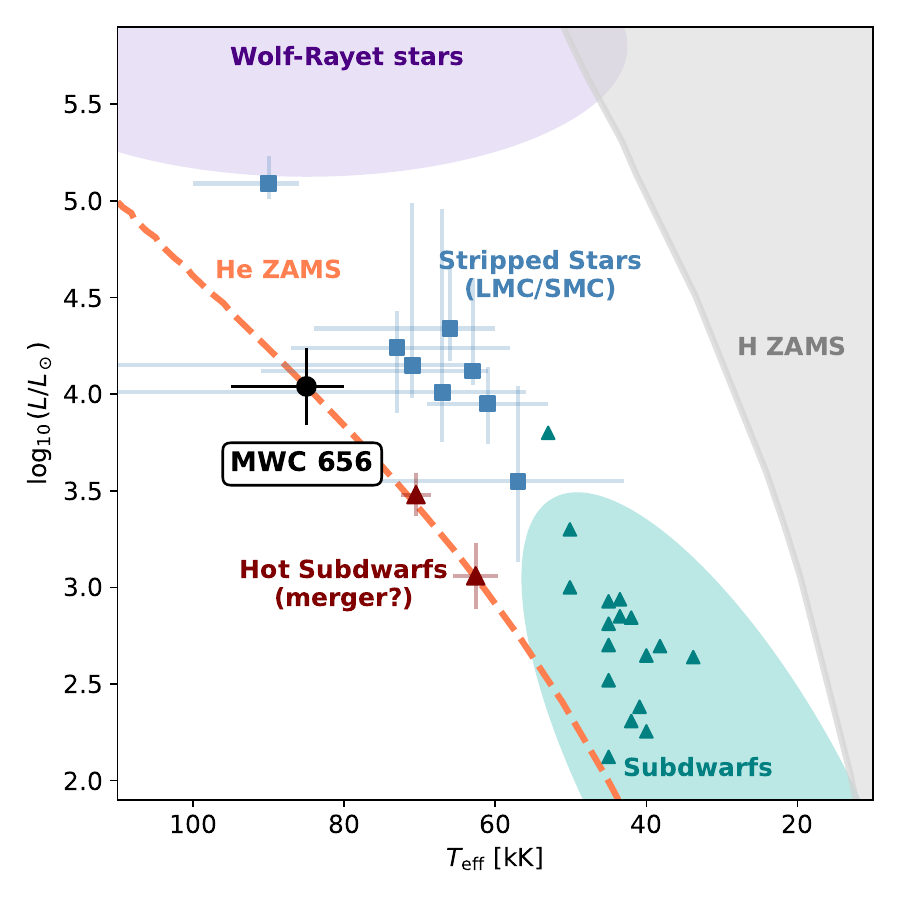}
    \caption{Hertzsprung-Russell diagram showing the inferred stellar parameters of the MWC~656 companion in comparison to known stripped stars. The lower-mass subdwarf sample \citep[green;][]{Wang+2021,Peters+2013,Peters+2016,Mourard+2015,Klement+2022}, two hot subdwarfs and merger product candidates \citep[maroon;][]{Krticka+2024}, intermediate-mass stripped stars in the Magellanic clouds \citep[blue;][]{Goetberg+2023}, and the companion in MWC~656 (black; $T_\mathrm{eff} = 85^{+10}_{-5}\,$kK, $\log L / \mathrm{L}_\odot = 4.0\pm 0.2$) occupy the region between the hydrogen ZAMS (grey) and the helium ZAMS (orange). Figure layout inspired by \citealt{Goetberg+2023}, their Fig.~8.} 
    \label{fig:stripped_star_comparison_hrd}
\end{figure} 

To place our results in context, we compare the inferred properties of the stripped star in MWC~656 to those of known stripped stars. These hot, compact helium stars are formed by envelope stripping in binary systems. Stripped stars span a wide mass range, from low-mass subdwarf B and O stars \citep[$\lesssim1.5\,\mathrm{M}_\odot$; e.g.][]{Heber2009,Mourard+2015,Klement+2022} to massive Wolf-Rayet stars \citep[$\gtrsim8\,\mathrm{M}_\odot$; e.g.][]{Shenar+2016,Shenar+2018,Shenar+2020b,SanderVink2020}, with several intermediate-mass cases recently identified in the Magellanic Clouds \citep{Ludwig+2025,Drout+2023,Ramachandran+2023,Ramachandran+2024}.

Stripped stars are often found with rapidly rotating main-sequence companions, believed to have spun up via accretion during envelope stripping. \citet{Wang+2021}, for example, identified Galactic Be+subdwarf candidates using HST UV spectroscopy, detecting hot subdwarf companions in ten of 13 systems. These systems provide a useful comparison to MWC~656 as they host rapidly rotating Be stars ($v \sin i \approx 200$--$400\,\mathrm{km\,s}^{-1}$) with effective temperatures of 19--27\,kK (spectral types B1--B3 with luminosity classes III--V), similar to MWC~656, accompanied by stripped subdwarfs.

In Fig.~\ref{fig:stripped_star_comparison_spectra} (top), we compare the UV spectrum of MWC~656 with these Be+subdwarf binaries (HST/STIS spectra; programme \texttt{15659}, PI: D. R. Gies). The Be stars dominate the flux in all systems, resulting in spectra that look overall similar to MWC~656. However, two differences stand out:
(a) MWC~656 shows much stronger wind lines in \ion{N}{v} than any of the Be+subdwarf binaries; and
(b) While the Be+subdwarf binaries show absorption near $1640\,\AA$ the corresponding feature in MWC~656 appears filled in by He~\textsc{ii} 1640\,\AA\ emission from the stripped companion (see also Appendix~\ref{app:disentangling}).

\citeauthor{Wang+2021} identified subdwarf companions through narrow cross-correlation peaks with 45\,kK O-type star templates. Applying the same method to MWC~656, we did not detect such a peak, suggesting a different type of stripped companion. In fact, while the \citeauthor{Wang+2021} subdwarfs have inferred temperatures around 45\,kK and luminosities $\log L / \mathrm{L}_\odot \approx 2.0$--$3.0$, our best-fit model for MWC~656 requires a much hotter ($\sim$85\,kK) and ten times more luminous stripped star.

Among the \citeauthor{Wang+2021} sample, only HD~55606 and HD~157832 show possible wind signatures and variability in \ion{N}{v}, though weaker than in MWC~656 (see Fig.~\ref{fig:stripped_star_comparison_sdO}). 
HD~55606 was first classified as a Be+subdwarf binary based on double-peaked \ion{He}{i} emission and weak \ion{He}{ii} absorption in the optical \citep{Chojnowski+2018}. \ion{He}{ii\,$\lambda4686$} in absorption in HD~55606, versus emission in MWC~656, could be due to the higher inferred mass and temperature of the companion in MWC~656 ($\sim$1.5\,M$_\odot$ compared to 0.92\,M$_\odot$).
In contrast, HD~157832 shows \ion{N}{v} variability but cross-correlation did not yield a companion detection by \citeauthor{Wang+2021}. Notably, HD~157832 is a known X-ray source \citep[$L_X = 1.3 \times 10^{32}$ erg\,s$^{-1}$;][]{LopesdeOliveira_Motch2011}, classifying it as a $\gamma$-Cas analogue. X-ray emission in $\gamma$-Cas-like systems has been proposed to arise from stripped star winds interacting with the Be-star disc \citep{Langer+2020b}, a scenario that may also apply to MWC~656.

Figure~\ref{fig:stripped_star_comparison_spectra} (bottom) also shows examples of accreting white dwarfs (cataclysmic variables). Their discs produce \ion{N}{v} and \ion{He}{ii} emission, often accompanied by strong C~\textsc{iv} \citep[e.g. 1RXS-J023238.8-371812 and ASAS0025+12][]{Pala+2022}, although there are exceptions such as HS0218+3229 \citep{Toloza+2023}, a white dwarf that accretes CNO-processed material and has weak C~\textsc{iv} emission. However, their broad line profiles differ markedly from the P Cygni-type features in MWC~656, which would require a disc wind or outflow.

Figure~\ref{fig:stripped_star_comparison_hrd} compares MWC~656 to known stripped stars in the Hertzsprung-Russell diagram.
The MWC~656 companion is hotter and more luminous than the \citeauthor{Wang+2021} subdwarfs, and hotter but similarly luminous compared to recently identified intermediate-mass stripped stars in the Large and Small Magellanic Clouds \citep{Drout+2023}. This aligns with expectations that lower-metallicity environments (like the Magellanic Clouds) lead to less efficient envelope stripping and cooler, only partially stripped stars \citep{Klencki+2022,DuttaKlencki2024}. The derived mass-loss rate is in good agreement with the \cite{Vink2017} prescription for stripped helium stars, as also found for partially stripped stars by \citet{Ramachandran+2024}, although we note that these studies focus on somewhat larger, cooler stars with lower surface gravities. In contrast, other detected stripped stars in the Magellanic Clouds are reported to have winds that are orders of magnitude weaker based on optical diagnostics \citep{Goetberg+2023}. We also include the helium-dominated merger products CD–46 8926 and CD–51 11879 \citep{Krticka+2024} in the figure, which are hot subdwarfs with $T_\mathrm{eff} > 60$\,kK and surface abundances pointing to a merger origin. Their luminosities and wind mass loss rates are an order of magnitude below those of MWC~656, yet they exhibit very similar UV wind features. In particular, CD–46 8926 shows a \ion{N}{v} doublet closely resembling that of the stripped companion in MWC~656 \citep[][their Fig.~5]{Krticka+2024}. For reference, we show the helium ZAMS for solar metallicity in Fig.~\ref{fig:stripped_star_comparison_hrd}, calculated by \cite{Picco+2024} using MESA evolutionary models \citep{Paxton+2011, Paxton+2013, Paxton+2015, Paxton+2018, Paxton+2019,Jermyn+2023}. The MWC~656 companion lies close to this theoretical boundary, although its inferred luminosity-to-mass ratio exceeds that expected for core-helium burning. The inferred luminosity implies an additional contribution, likely due to He-shell burning, suggesting that the companion may already have evolved beyond the core-He-burning phase. Detailed binary evolution models will be required to test this interpretation.

\subsection{Caveats of the composite model}
\label{sec:discussion_caveats}

A notable feature of the UV data is an apparent Doppler shift in the high-ionisation wind lines. In three of the four epochs, the lines show no significant shift, but at $\Phi=0.4$ the red edge of the absorption component, clearly seen, for example, in \ion{C}{iv} (Fig.~\ref{fig:spec_uv_zoom}), appears redshifted by $\sim$80\,km\,s$^{-1}$. The measurement of RVs in wind lines is inherently difficult due to their complex and variable profiles, and the situation could be further complicated here by interstellar absorption that overlaps with the wind features. Nevertheless, the observed shift cannot easily be explained by the orbital motion of either star: The amplitude is too small for the companion, the sign is opposite to the expected motion at this phase, and the shift is too large to be attributed to the Be star. This suggests that the high-ionisation line-forming region does not simply trace the orbital motion of either component. Redshifted absorption in UV wind lines has been reported in colliding-wind binaries, where wind-wind interaction regions can lead to orbital phase-dependent features \citep{Stevens1993}. The applicability of this interpretation to MWC~656 is uncertain given the weak wind of the Be star. Alternatively, the interaction between the wind of the stripped star and the disc and/or the wind of the Be star could generate velocity components decoupled from orbital motion and could also account for the observed X-ray emission of the system \citep{Langer+2020b}. 

Additional evidence of wind--disc interaction comes from the transient appearance of metastable \ion{He}{i$^*$} absorption lines, which are known tracers of outer-disc irradiation. In four ARCES spectra near quadrature ($\Phi = 0.36$--0.46), we detect sharp absorption in \ion{He}{i\,$\lambda\lambda3889,5017$}, consistent with heating of the Be star’s disc by the hot companion. Further observations are needed to test this scenario; in particular, a UV spectrum obtained near inferior conjunction -- when the companion passes in front of the Be star -- would probe the hemisphere of the stripped star facing away from the Be star’s radiation, where wind-disc interaction is expected to be weaker.

Another possible signature of interaction is the deep, narrow, near-rest frame absorption component observed in \ion{N}{v} (Fig.~\ref{fig:spec_uv_zoom}). Its depth is too large to be explained by a stripped companion contributing only $\lesssim$10--15\% of the total flux, instead requiring a substantial fraction of the Be-star light to be absorbed by low-velocity, ionised material along the line of sight. This material may be associated with the Be star that is irradiated by the hot companion. An origin intrinsic to the Be star itself appears unlikely, as \ion{N}{v} is not expected for a mid-type Be star; this is also illustrated by comparison with other Be+subdwarf systems (Fig.~\ref{fig:stripped_star_comparison_spectra}), none of which show similarly strong, narrow \ion{N}{v} absorption, with only a partial exception in the earlier-type system 60~Cyg. Moreover, the absorption does not follow the orbital motion of the Be star.
Although the physical origin of this absorption component remains uncertain, it does not affect our main conclusion that a hot, stripped, luminous companion provides the most plausible explanation for the underlying \ion{N}{v} P-Cygni profile, including its emission and blueshifted absorption trough.

\section{Summary and conclusions}
\label{sec:conclusion}

MWC~656 has been proposed as the first quiescent Be+BH binary in the Milky Way. In this study, we revisit this interpretation based on new optical and UV spectroscopic observations and argue that the presence of a BH companion can be ruled out for this system.

Using new high-resolution optical spectra, we revised the orbital solution of the system. Our updated binary parameters yield minimum dynamical masses of $5.13\,$M$_\odot$ for the Be star and $1.03$\,$M_\odot$ for its companion. A detailed spectroscopic analysis of the Be star allows us to constrain its effective temperature, luminosity, and mass more robustly, resulting in an estimated spectroscopic mass of $7.4 \pm 2.7\,\mathrm{M}_\odot$. Together with dynamical constraints, this implies a companion mass of approximately $1.48^{+0.55}_{-0.46}\,\mathrm{M}_\odot$.

This mass range is significantly below the canonical minimum mass for BHs, and mostly above the Chandrasekhar limit for white dwarfs, placing the companion in the neutron star or intermediate-mass stripped star regime. These findings are consistent with earlier concerns \citepalias[][]{Rivinius+2024,Janssens+2023} that the companion's mass is too low to be a BH.

To further investigate the nature of the companion, we obtained multi-epoch HST/STIS FUV spectra of MWC~656. These reveal high-ionisation wind features, such as N~\textsc{v} lines with P Cygni-like profiles, which cannot originate in the Be star. The line strengths vary with orbital phase, increasing near quadrature and becoming weaker near conjunction when the companion is behind the Be star. This behaviour suggests partial obscuration of the line-forming region and supports an origin in the hot wind of a luminous companion.

We considered alternative scenarios in which the companion is a compact object accreting from the Be star's wind. However, we find that the level of X-ray emission required to power the observed wind features is inconsistent with the observed X-ray luminosity of the system, and such models would predict additional spectral features, such as strong C~\textsc{iv} emission, which are not observed.

Instead, we favour an interpretation in which the companion is a hot, stripped star with a radiatively driven wind. We modelled the combined UV and optical spectrum using a composite of a Be star and a helium-enriched companion, successfully reproducing the key UV wind lines and He~\textsc{ii} emission in the spectrum. The best-fitting model suggests a companion with a temperature of $85^{+10}_{-5}$\,kK, and luminosity of approximately $10^4$\,L$_\odot$ (see Fig.~\ref{fig:composite_spectral_fit} and Table~\ref{tab:stellar_parameters}). With that, MWC~656 is another example in the growing sample of detected Be + stripped star binaries. The inferred temperature of the companion is higher than that of previously known Be + stripped star binaries of comparable luminosity in the Magellanic Clouds. This could be a consequence of less efficient envelope stripping at lower metallicities.

Although the nature of RV variability in the wind lines and the origin of the deep and narrow absorption component in \ion{N}{v} remain uncertain, the presence of a hot stripped companion provides a coherent explanation for most of the UV features. This interpretation not only rules out the BH scenario, but highlights MWC~656 as a valuable system for studying the population of intermediate-mass stripped stars. Depending on the precise mass of the Be star, the system may eventually evolve into a double white dwarf or neutron star binary, making it a compelling target for future population synthesis and evolutionary studies.

\section*{Data availability}
The newly calculated PoWR model spectra for the Be star and the companion will be provided as supplementary material on \href{https://doi.org/10.5281/zenodo.18315141}{Zenodo}. 
The HST UV spectra used in this work are publicly accessible and can be obtained from
\href{http://dx.doi.org/10.17909/cfms-dg96}{MAST}. Optical TRES and HIRES spectra are only available in electronic form at the CDS via anonymous ftp to cdsarc.u-strasbg.fr (130.79.128.5) or via http://cdsweb.u-strasbg.fr/cgi-bin/qcat?J/A+A/. 

\begin{acknowledgements}
The authors sincerely thank the anonymous referee for their thorough review and thoughtful and constructive comments.
The authors thank Howard Isaacson and Andrew W. Howard for their help in obtaining five HIRES optical spectra, which were used in the orbital analysis, and Kevin Burdge and Charlie Conroy for their help in obtaining the HST spectra. J. M\"uller-Horn thanks Hans-Walter Rix, Sahar Shahaf, Pablo Marchant, Annachiara Picco, Anna F. Pala, and Dominic Bowman for helpful discussions and suggestions.  J. M\"uller-Horn acknowledges support from the European Research Council for the ERC Advanced Grant [101054731]. V. Ramachandran and A.A.C. Sander are supported by the German
\textit{Deut\-sche For\-schungs\-ge\-mein\-schaft, DFG\/} in the form of an Emmy Noether Research Group -- Project-ID 445674056 (SA4064/1-1, PI Sander) and acknowledge financial support by the Federal Ministry for Economic Affairs and Energy (BMWE) via the Deutsches Zentrum f\"ur Luft- und Raumfahrt (DLR) grant 50 OR 2509 (PI Sander). V. Ramachandran, A.A.C. Sander, and E.C. Sch{\"o}sser further acknowledge financial support by the BMWE via the DLR grant 50 OR 2306 (PI Ramachandran/Sander).
This project was co-funded by the European Union (Project 101183150 - OCEANS).
This work has made use of data from the European Space Agency (ESA) mission \textit{Gaia} (https://www.cosmos.esa.int/gaia), processed by the \textit{Gaia} Data Processing and Analysis Consortium (DPAC, https://www.cosmos.esa.int/web/gaia/dpac/consortium). Funding for the DPAC has been provided by national institutions, in particular the institutions participating in the \textit{Gaia} Multilateral Agreement.
TS acknowledges support from the Israel Science Foundation (ISF) under grant number 0603225041 and from the European Research Council (ERC) under the European Union's Horizon 2020 research and innovation program (grant agreement 101164755/METAL).

\end{acknowledgements}

%
   \bibliographystyle{aa} 
   \bibliography{references.bib} 
%

\appendix

\section{Stellar abundances}
\label{app:abundances}

The abundances adopted in the composite spectral analysis are given in Table~\ref{tab:stellar_abundances}. For the Be star, C, N, and O mass fractions were adjusted to match diagnostic lines in the optical and UV spectra. Here, we focussed in particular on \ion{N}{ii\,$\lambda\lambda 1598$--$99, 3995, 4428$--$4433, 4241$--$42$},  \ion{O}{ii\,$\lambda\lambda
 4007, 4085, 4649$}, \ion{O}{iii\,$\lambda\lambda
1768, 1410$--$12$},
 \ion{C}{ii\,$\lambda\lambda
4267$} and \ion{C}{iii\,$\lambda\lambda
1909, 2298, 1247, 4647$--$4651$}. For the stripped companion, which lacks spectroscopic constraints, we adopted fixed values guided by analyses of known stripped stars. For other elements, we assumed solar abundances as derived
by \cite{Asplund+2009} and implemented in \cite{Hainich+2019}.

\begin{table}[t]
        \caption{Stellar abundances assumed for MWC~656 in the composite spectroscopic analysis.}
        \label{tab:stellar_abundances}
        \centering
        \renewcommand{\arraystretch}{1.6}
        \begin{tabular}{lcc}
                \hline
                \hline
                \vspace{0.1cm}
                &       Be star &       Stripped star \\
                \hline
                $X_{\rm H}$ (mass fr.) & 0.737 & 0.2 \\
                $X_{\rm He}$ (mass fr.) & 0.26 & 0.79 \\
        $X_{\rm C}$ (mass fr.) & 0.0008 & 0.0004 \\
                $X_{\rm N}$ (mass fr.) & 0.002 & 0.003 \\
        $X_{\rm O}$ (mass fr.) & 0.003 & 0.002 \\
                \hline
        \end{tabular}
\end{table}

\section{Spectral disentangling in the UV}
\label{app:disentangling}
We performed spectral disentangling of the UV spectra using the same approach as for the optical spectra (see Sect.~\ref{sec:disentangling}), aiming to separate the contributions of the Be star and its companion. Here, we focussed on the wavelength range 1635--1650\,Å, where there is clearly visible phase-dependent variability (see Fig.~\ref{fig:spec_uv_zoom}).

We did not disentangle in regions with high-ionisation wind lines, such as N~\textsc{v}. This is because these lines exhibit strong line profile variability that violates the key assumption of time-invariant component spectra underlying the disentangling technique.

\begin{figure}
    \centering
    \begin{minipage}[t]{\columnwidth}
        \centering
        \includegraphics[width=\textwidth]{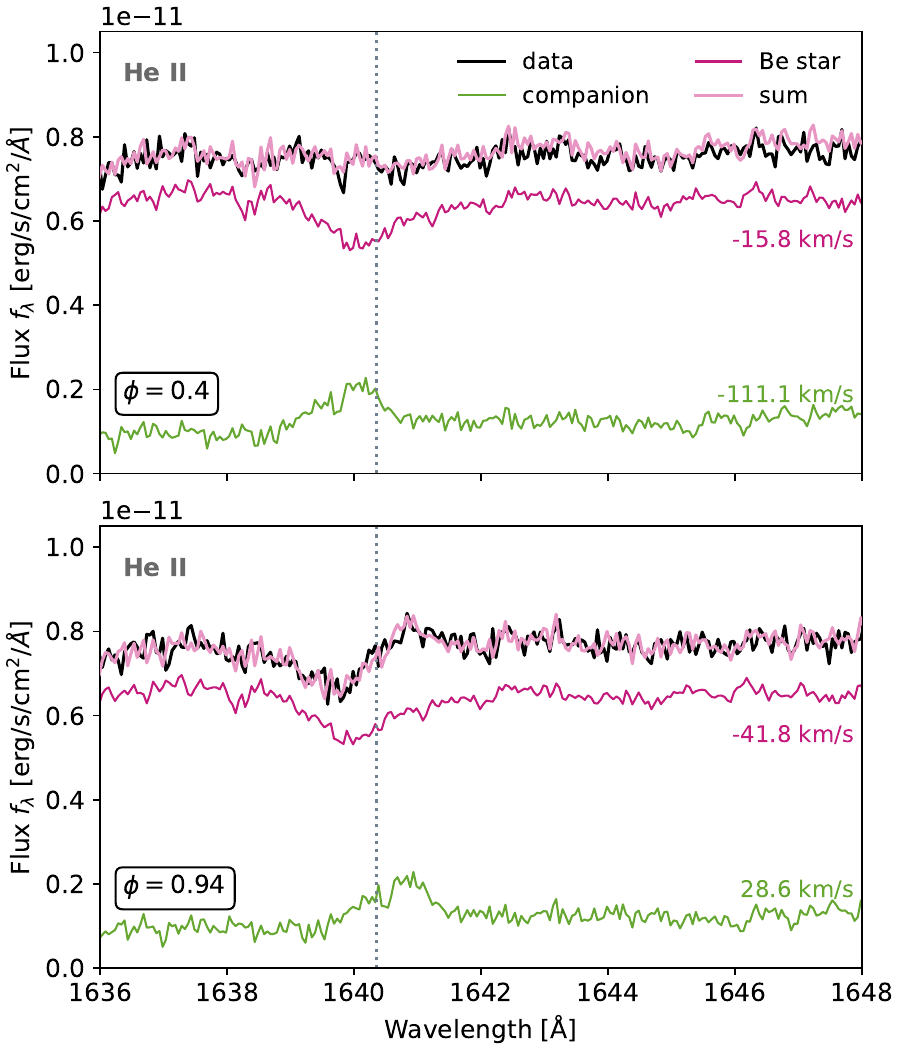}
    \end{minipage} 
    \caption{Disentangled HST UV spectra of MWC~656 around the He~\textsc{ii} 1640\,Å line at orbital phases close to quadrature. The disentangled and Doppler-shifted spectra of the Be star (pink) and the companion (green) sum to the combined model spectrum (light pink), which closely reproduces the observed spectra (black). The disentangled spectra show an emission component for the companion that is moving in anti-phase with the Be star.} 
    \label{fig:disentangling_UV}
\end{figure}

The resulting disentangled spectra of the Be star and companion are shown in Fig.~\ref{fig:disentangling_UV} at orbital phases near quadrature. For the Be star, we recover a broad absorption feature similar to what is observed in other Be stars of similar temperature (see Fig.~\ref{fig:stripped_star_comparison_spectra}). For the companion, the disentangled spectrum reveals an emission feature, most likely corresponding to He~\textsc{ii} 1640\,Å. This is consistent with our spectral model of the companion as a hot stripped star, which predicts He~\textsc{ii} emission (see Fig.~\ref{fig:composite_spectral_fit}), and agrees with the observed He~\textsc{ii} 4686\,Å emission in the optical spectra. The emission does not show a double-peaked profile; however, the limited number of available UV spectra (only four epochs) means that the shape and strength of the extracted component spectra remain uncertain.

\section{Stellar parameters of the Be star}
\label{appendix:stellar_params}

\begin{figure*}[h!]
    \centering
    \includegraphics[width=0.6\textwidth, angle=90]{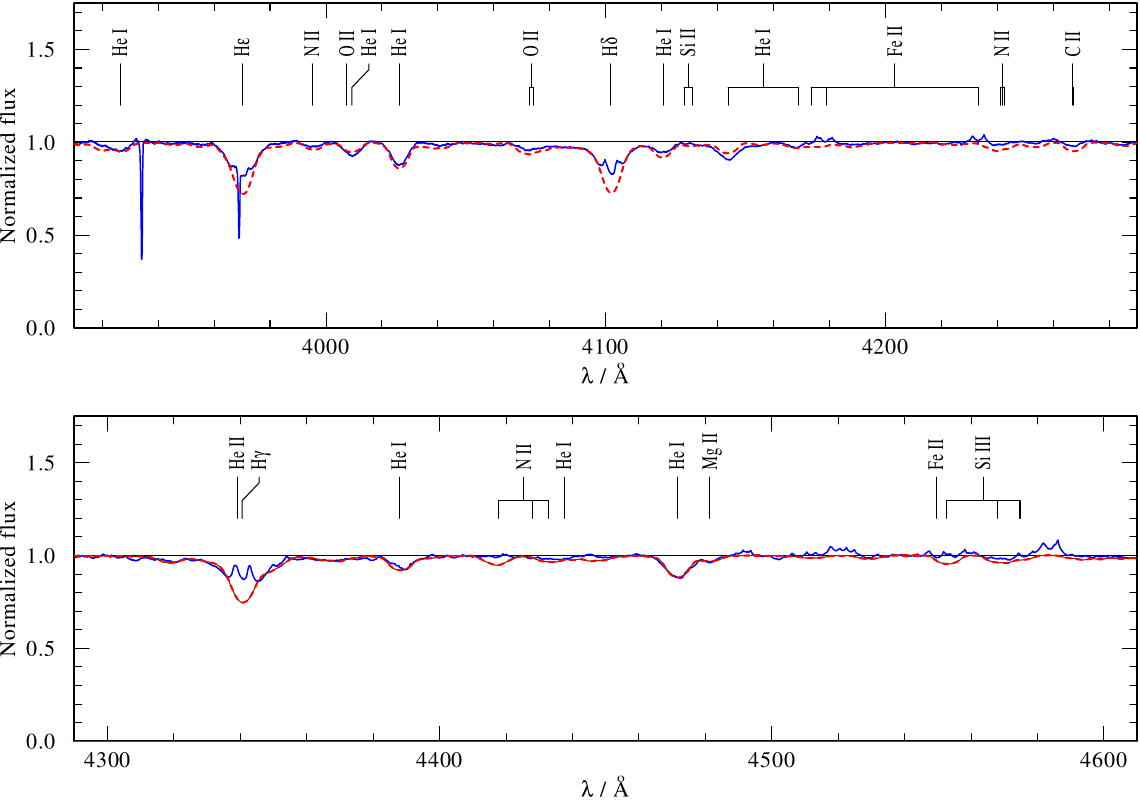}
    \caption{Selected diagnostic wavelength regions of the observed optical spectrum of MWC~656 (blue) compared to the best-fit PoWR stellar atmosphere model for the Be star (red) with $T_* = 21$\,kK, $\log g_* = 3.4$, and $v\sin i = 300$\,km\,s$^{-1}$. The narrow absorption lines result from interstellar absorption.}
    \label{fig:MWC656_opt_zoom}
\end{figure*}

Figure~\ref{fig:MWC656_opt_zoom} shows a close-up comparison between the observed optical spectrum of MWC~656 and the best-fit model for the Be star. Highlighted are several diagnostic lines that were used to infer the stellar parameters, including effective temperature, surface gravity, and projected rotational velocity.

\section{Observation overview}
\label{appendix:obs_overview}

\setlength{\extrarowheight}{3pt}
\begin{table*}[t]
\caption{Summary of spectroscopic observations.}\label{tab:obs_overview}
\centering
\begin{tabularx}{0.85\textwidth}{c c c c c c c c}
\toprule
Instrument & MJD & S/N & Orbital Phase & RV$_1$(He\,\textsc{i}) & $\sigma_{\mathrm{RV1}}$ & RV$_2$(He\,\textsc{ii}) & $\sigma_{\mathrm{RV2}}$ \\
 & (d) & & & (km\,s$^{-1}$) & (km\,s$^{-1}$) & (km\,s$^{-1}$) & (km\,s$^{-1}$) \\
\midrule
\textit{Optical} & & & & & \\
\midrule 
ARCES & 57643.23 & 150.8 & {0.36} & {$-28.28$} & {3.87} & {$-117.01$} & {5.77} \\
ARCES & 58000.29 & 103.4 & {0.41} & {$-2.80$} & {4.17} & {$-106.42$} & {5.45} \\
ARCES & 59540.19 & 173.7 & {0.49} & {$-18.51$} & {4.04} & {$-78.76$} & {4.43} \\
ARCES & 59545.15 & 148.7 & {0.56} & {$-21.87$} & {3.72} & {$-23.23$} & {4.20} \\
ARCES & 58085.15 & 153.2 & {0.85} & {$-47.81$} & {3.90} & {$+48.89$} & {3.90} \\
ARCES & 59007.38 & 128.5 & {0.47} & {$-28.05$} & {3.93} & {$-98.30$} & {4.04} \\
ARCES & 59015.41 & 136.3 & {0.60} & {$-35.28$} & {4.18} & {$-22.82$} & {3.36} \\
ARCES & 59186.22 & 128.9 & {0.50} & {$-25.72$} & {4.42} & {$-84.62$} & {3.89} \\
ARCES & 59214.10 & 105.5 & {0.97} & {$-40.37$} & {6.57} & {$+8.76$} & {1.68} \\
ARCES & 59216.09 & 202.2 & {0.00} & {$-35.41$} & {3.81} & {$+2.38$} & {1.27} \\
ARCES & 59536.22 & 167.3 & {0.43} & {$-6.26$} & {3.34} & {$-102.43$} & {5.41} \\
ARCES & 59538.04 & 207.4 & {0.46} & {$-16.84$} & {3.50} & {$-97.20$} & {3.94} \\
\addlinespace
ESPaDOnS & 57201.55 & 373.2 & {0.88} & {$-30.34$} & {3.04} & {$+39.37$} & {2.24} \\
ESPaDOnS & 57208.57 & 379.9 & {1.00} & {$-29.66$} & {3.01} & {$+1.30$} & {0.94} \\
ESPaDOnS & 57227.57 & 217.8 & {0.32} & {$-7.88$} & {2.80} & {$-121.30$} & {3.32} \\
\addlinespace
HIRES & 60189.36 & 198.2 & {0.49} & {$-47.04$} & {3.68} & {$-74.52$} & {2.41} \\
HIRES & 60213.36 & 178.7 & {0.90} & {$-58.78$} & {4.11} & {$+41.34$} & {2.25} \\
HIRES & 60271.19 & 176.8 & {0.88} & {$-54.33$} & {3.65} & {$+43.76$} & {2.09} \\
HIRES & 60274.30 & 186.3 & {0.93} & {$-54.42$} & {3.61} & {$+30.38$} & {1.81} \\
HIRES & 60275.37 & 144.9 & {0.95} & {$-53.00$} & {3.36} & {$+23.00$} & {1.63} \\
\addlinespace
TRES & 59857.26 & 39.3 & {0.86} & {$-20.05$} & {9.98} & {$+38.65$} & {5.88} \\
TRES & 59864.28 & 50.0 & {0.98} & {$-46.54$} & {7.80} & {$+11.90$} & {3.33} \\
TRES & 59870.22 & 59.5 & {0.08} & {$-20.82$} & {8.18} & {$-33.32$} & {5.21} \\
TRES & 59879.29 & 70.9 & {0.24} & {$-11.95$} & {7.31} & {$-93.06$} & {4.65} \\
TRES & 59885.26 & 66.7 & {0.34} & {$-20.42$} & {6.68} & {$-119.37$} & {4.67} \\
TRES & 59891.19 & 50.3 & {0.44} & {$-6.78$} & {7.89} & {$-94.91$} & {5.67} \\
TRES & 59900.26 & 25.0 & {0.59} & {$-24.84$} & {11.85} & {$-6.95$} & {8.48} \\
TRES & 59912.17 & 29.8 & {0.79} & {$-49.20$} & {11.65} & {$+42.08$} & {8.19} \\
TRES & 59922.22 & 42.2 & {0.96} & {$-19.54$} & {10.59} & {$+10.83$} & {6.15} \\
\midrule
\textit{Ultraviolet} & & & & & & & \\
\midrule
HST/STIS & 59992.7 & 40 & {0.16} & \multicolumn{4}{c}{---} \\
HST/STIS & 59920.6 & 46 & {0.94} & \multicolumn{4}{c}{---} \\
HST/STIS & 59947.7 & 44 & {0.40} & \multicolumn{4}{c}{---} \\
HST/STIS & 60347.3 & 45 & {0.17} & \multicolumn{4}{c}{---} \\
HST/COS & 59992.7 & 39 & {0.16} & \multicolumn{4}{c}{---} \\
HST/STIS (FUV) & 59934.0 & 10 & {0.17} & \multicolumn{4}{c}{---} \\
\bottomrule
\end{tabularx}
\tablefoot{Columns list the instrument, observation date (MJD), S/N, and orbital phase for each spectrum, measured RVs and uncertainties for the Be star (from He~I absorption) and the companion (from He~\textsc{ii} emission) for the optical spectra. To convert the relative RVs of the companion to absolute values (Fig.~\ref{fig:orbit_fit}), the constant velocity offset $(v_z - v_{\mathrm{rel.}, 2}) = 3.2 \,$km\,s$^{-1}$ needs to be added.}
\end{table*}
\setlength{\extrarowheight}{0pt}

In Table~\ref{tab:obs_overview} we provide an overview of the optical observations including measured RVs and S/N estimates. 
\section{Temperature of the stripped star}
\label{app:stripped_star}

\begin{figure*}[ht]
    \centering
    \includegraphics[width=0.9\textwidth]{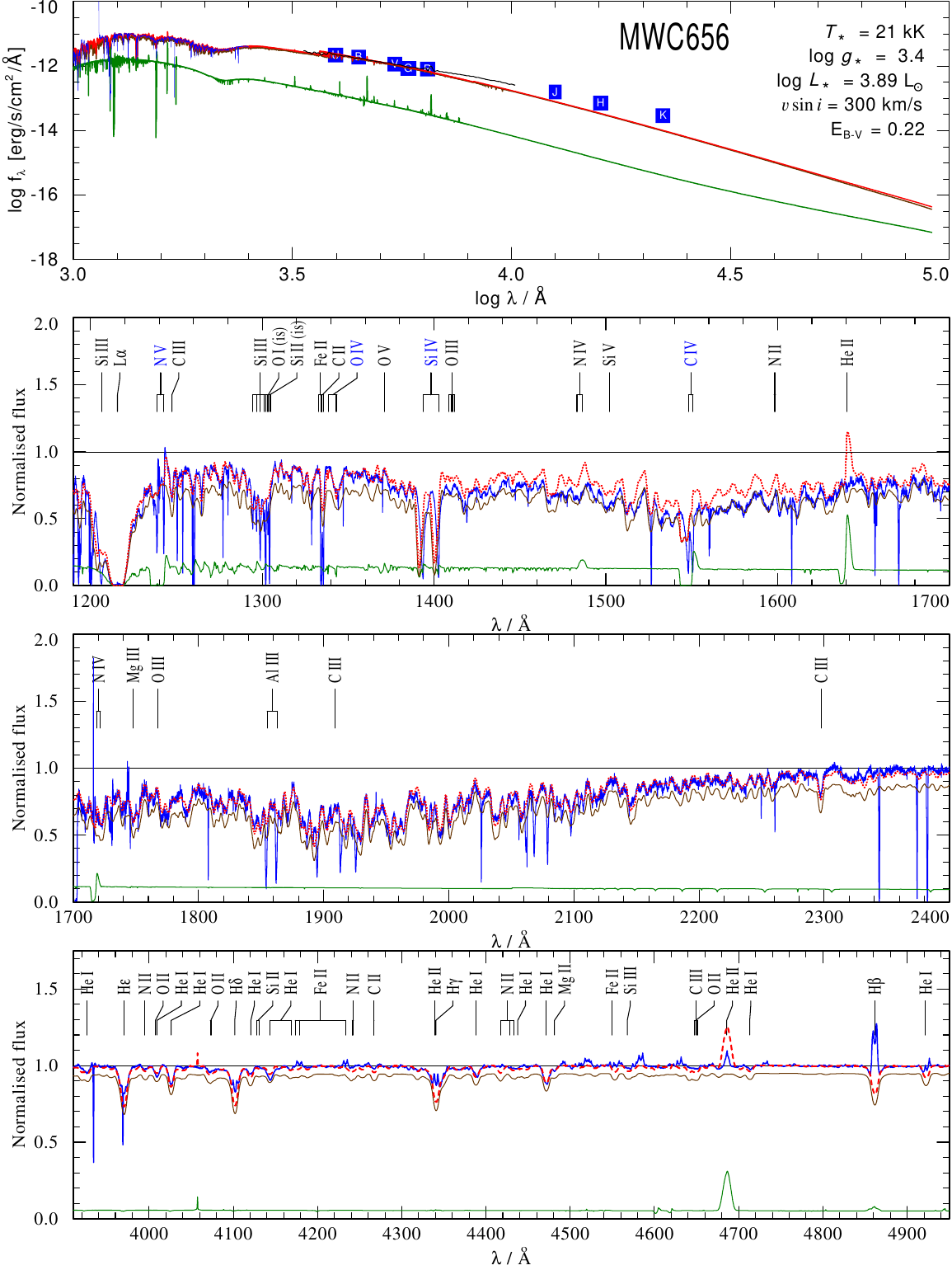}
    \caption{Observed spectra and photometry of MWC~656 (blue) compared to composite model fits (red). The models combine the rapidly rotating Be star (brown) with a stripped companion (green). The example shown here assumes a cooler stripped star ($T_\ast = 65$\,kK) than in the fiducial best-fit model ($T_\ast = 85$\,kK). Cooler models strengthen \ion{C}{iv} and over-predict the emission in \ion{He}{ii}.} 
    \label{fig:composite_spectral_fit_65kK}
\end{figure*}

\begin{figure*}[ht]
    \centering
    \includegraphics[width=0.9\textwidth]{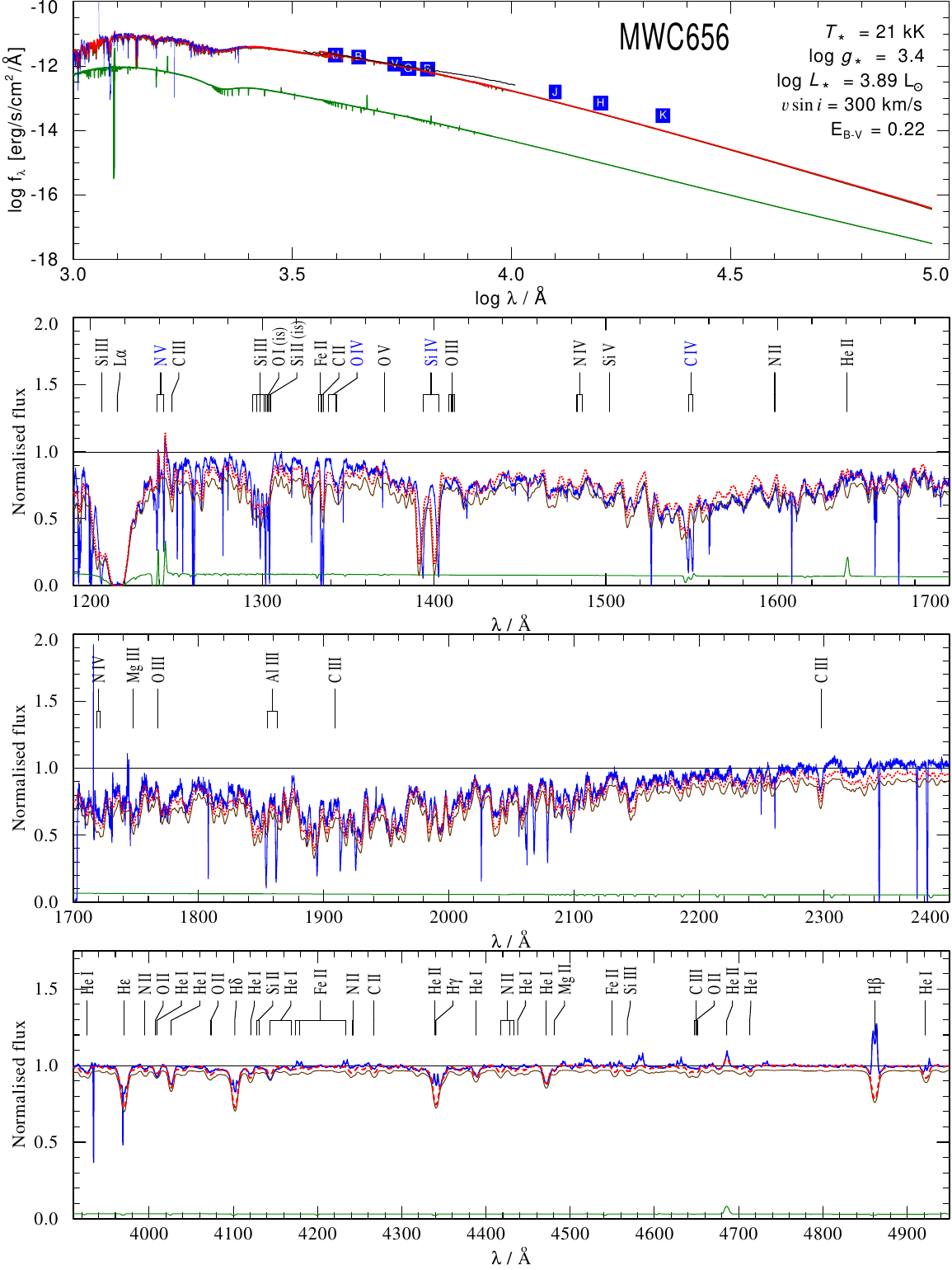}
    \caption{Observed spectra and photometry of MWC~656 (blue) compared to composite model fits (red). The models combine the rapidly rotating Be star (brown) with a stripped companion (green). The example shown here assumes a hotter stripped star ($T_\ast = 95$\,kK) than in the fiducial best-fit model ($T_\ast = 85$\,kK). Hotter models have a steeper SED and comparably lower flux contribution in the optical, visible in the \ion{He}{ii\,$\lambda4686$} line.} 
    \label{fig:composite_spectral_fit_95kK}
\end{figure*}

To constrain the temperature of the stripped star, we fitted composite models with companion effective temperatures ranging from 65 to 95\,kK. Figures~\ref{fig:composite_spectral_fit_65kK} and \ref{fig:composite_spectral_fit_95kK} compare these models with the observations. Lower-temperature models ($\sim$65\,kK) produce stronger \ion{C}{iv} features, underestimate the observed \ion{N}{v} doublet, and overpredict \ion{He}{ii} emission lines, while hotter models ($\gtrsim$95\,kK) steepen the SED and underestimate the optical contribution. Our best-fit solution is achieved at $T_\ast \approx 85$\,kK (Sect.~\ref{sec:model}), balancing these aspects.


\section{Circumsecondary accretion disc}
\label{app:circumsecondary_disc}
In our best-fit composite model, the \ion{He}{ii} emission in both the optical and UV spectra is naturally explained by the atmosphere and wind of the stripped star. Previous studies, however, attributed the slightly double-peaked \ion{He}{ii$\,\lambda4686$} profile to a Keplerian accretion disc around the companion \citepalias{Rivinius+2024}. These interpretations are not mutually exclusive: an accretion disc around a stripped companion remains plausible. In our analysis, the optical \ion{He}{ii} line strength provides a constraint on the stripped star’s temperature; if part of the emission originates in a disc rather than the stellar wind, the required stellar temperature would shift to higher values (see Sect.~\ref{sec:model}). However, we note that a double-peaked profile is not unique to Keplerian discs. Hydrodynamically consistent atmosphere models for stripped stars also predict double-peaked \ion{He}{ii} profiles \citep{Sabhahit2025}. 

The geometry of the system supports the possibility of mass transfer. From the H$\alpha$ equivalent width and using the algorithm of \citet{Grundstrom_Gies2006}, we estimated the Be star’s disc half-width at half-maximum emission radius to be $\simeq8.4\,R_\ast$, comparable to its Roche-lobe radius ($\simeq10\,R_\ast$). This suggests that the outer Be disc can extend far enough to lose gas towards the secondary. Numerical simulations of Be + stripped-star systems indicate that such a transfer can form Keplerian accretion discs around the companion \citep[e.g.][]{Panoglou+2016,Rubio+2025}, and predict X-ray luminosities consistent with those observed in MWC~656 \citep{Rast+2025}.
Assuming that the \ion{He}{ii} line-peak separation traces Keplerian motion, we infer an outer radius of $\sim34\,R_\odot$ ($\simeq0.26\,a$, where $a$ is the orbital semi-major axis), similar to the Roche-lobe radius of the companion. 
Such a disc around the companion could contribute to the UV flux and host very hot plasma ($T\sim10^5$\,K; \citealt{Peters_Polidan1984}), potentially obscuring part of the companion's flux and producing redshifted \ion{N}{v} absorption, as seen in HR~2142 \citep{Peters+2016}. However, the observed \ion{He}{ii} emission lines are narrower than expected for a Keplerian disc around a high-gravity ($\log g_* \simeq 5$) star -- the inferred inner-disc rotation speed would exceed the measured width by approximately a factor of two. Moreover, a circumsecondary disc cannot reproduce the P Cygni-like \ion{N}{v} profile, arguing against the disc as the dominant source of the optical \ion{He}{ii} emission.

\end{document}